\documentclass[12pt]{iopart}
\usepackage{graphicx}
\usepackage{cite}

\begin{document}

\title[Quasispecies dynamics on a network of interacting genotypes and
idiotypes]
{Quasispecies dynamics on a network of interacting genotypes and idiotypes:
applications to autoimmunity and immunodeficiency}

\author{Valmir C. Barbosa$^1$, Raul Donangelo$^{2,3}$ and Sergio R. Souza$^{2,4}$}

\address{$^1$ Programa de Engenharia de Sistemas e Computa\c c\~ao, COPPE,
Universidade Federal do Rio de Janeiro,
Caixa Postal 68511, 21941-972 Rio de Janeiro - RJ, Brazil}
\address{$^2$ Instituto de F\'\i sica,
Universidade Federal do Rio de Janeiro,
Caixa Postal 68528, 21941-972 Rio de Janeiro - RJ, Brazil}
\address{$^3$ Instituto de F\'\i sica, Facultad de Ingenier\'\i a,
Universidad de la Rep\'ublica,
Julio Herrera y Reissig 565, 11.300 Montevideo, Uruguay}
\address{$^4$ Instituto de F\'\i sica,
Universidade Federal do Rio Grande do Sul,
Caixa Postal 15051, 91501-970 Porto Alegre - RS, Brazil}

\ead{valmir@cos.ufrj.br}

\begin{abstract}
In spite of their many facets, the phenomena of autoimmunity and
immunodeficiency seem to be related to each other through the subtle links
connecting retroviral mutation and action to immune response and adaptation. In
a previous work, we introduced a network model of how a set of interrelated
genotypes (called a quasispecies, in the stationary state) and a set of
interrelated idiotypes (an idiotypic network) interact. That model, which does
not cover the case of a retroviral quasispecies, was instrumental for the study
of quasispecies survival when confronting the immune system and led to the
conclusion that, unlike what happens when a quasispecies is left to evolve by
itself, letting genotypes mutate too infrequently leads to the destruction of
the quasispecies. Here we extend that genotype-idiotype interaction model by the
addition of a further parameter ($\nu$) to account for the action of
retroviruses (i.e., the destruction of idiotypes by genotypes). We give
simulation results within a suitable parameter niche, highlighting the issues of
quasispecies survival and of the onset of autoimmunity through the appearance of
the so-called pathogenic idiotypes. Our main findings refer to how $\nu$ and
$\lambda$, a parameter describing the rate at which idiotypes get stimulated,
relate to each other. While for $\nu>\lambda$ the quasispecies survives at the
expense of weakening the immune system significantly or even destroying it, for
$\nu<\lambda$ the fittest genotypes of the quasispecies become mimicked inside
the immune system as pathogenic idiotypes. The latter is in agreement with the
current understanding of the HIV quasispecies.

\end{abstract}

\maketitle

\section{Introduction}
\label{sec:intro}

Since its introduction more than four decades ago, the term quasispecies has
been used to refer to the stationary state of a set of interrelated genotypes
that mutate frequently into one another without recombination \cite{e71,es77}.
This mutational dynamics is based on fitnesses that do not depend on genotype
abundance and has led the concept of a quasispecies to be used in the modeling
of interacting complex entities such as prebiotic molecules and RNA viruses,
for example \cite{e71,es77,d09,la10,mlcgm10}. With our recent introduction of
new elements aiming to make the theory biologically more plausible \cite{bds12},
it has become possible to analyze the quasispecies dynamics as a process taking
place on a network of genotypes, clarifying key aspects not only of the dynamics
itself but also of the quasispecies' eventual survival or demise. The most
crucial element of this network-based formulation is a probability parameter,
$p$, used not only to create the random graph that underlies all mutations but
also to regulate the mutational dynamics itself. We have found, in conformity
with the theory's basic assertions, that very low values of $p$ practically
guarantee the survival of the quasispecies and that, by contrast, progressively
higher values of $p$ (which let not only each genotype mutate into more of the
others but also mutate more frequently) gradually lead the quasispecies to its
destruction.

In a further recent work \cite{bds15}, we demonstrated moreover that such a
network of genotypes can be used to substitute for the isolated, essentially
noninteracting antigens traditionally employed as external inputs to models of
the immune system. Doing this lets the immune-system model in use be presented
with a great variety of interrelated genotypes, much as occurs in the case of
several viral infections, therefore elevating the modeling effort to a higher
level of plausibility. In the model we use in \cite{bds15}, the immune system is
represented by a network as well, in the spirit of the idiotypic network
introduced, interestingly, at about the same time as the notion of a
quasispecies \cite{j74}. This is a network of idiotypes, that is, a network
encompassing the great variety of immune-stimulating receptors that are present
not only in antigens but also in those elements of the immune system (such as
molecules and cells) whose task is precisely to confront and eliminate antigens.
This latter observation lies at the heart of the idiotypic-network theory of
immunity: unlike its main contender, the antigen-centered theory of clonal
selection \cite{b59,f95}, it postulates the existence of a complex dynamics of
idiotypes even in the absence of antigens (that is, when the immune system is in
its innate state). Owing to this fundamental distinction, important aspects of
the idiotypic-network theory have found their way into several models of the
immune system \cite{fabc04,b07,mkbmqtcb11}.

Our own model of the immune system in \cite{bds15} is similar to that of a
quasispecies in \cite{bds12}. It is based on another probability parameter, $r$,
used both to create a random graph to underlie idiotype interactions and to
regulate the inter-idiotype stimulations on which such interactions are based.
The model becomes complete when the $p$-based network of genotypes and the
$r$-based network of idiotypes are joined together to represent the interaction
of the quasispecies with the immune system. This is achieved by creating further
connections, now solely directed from genotypes toward idiotypes but still based
on the probability $r$. Depending on two further parameters (the rates $\lambda$
and $\mu$, to be revisited later), the complete model allows for a dynamics of
interactions in which genotypes both mutate into one another and stimulate the
idiotypes while the latter stimulate one another. Stimulation in one direction
entails recognition in the other, so whenever stimulation is being effected the
stimulated idiotype reacts to reduce the stimulator's abundance. This, in turn,
opens the way both for the quasispecies to be destroyed when the immune system
has the upper hand and for the immune system to reorganize itself. Contrasting
with the results for the isolated network of genotypes in \cite{bds12}, now a
minimum value of $p$ exists below which the quasispecies no longer survives.

Here we extend the work reported in \cite{bds15} by concentrating on two key
aspects that we had left untouched. The first one is that of autoimmunity, that
is, the biological condition in which the immune system turns against the body
that it should be protecting. The onset of autoimmunity has been linked to how
the immune system gets organized in early life, that is, during the initial
transitions out of its innate state \cite{yfkbm15}. Notwithstanding this, that
such a condition should be able to arise can be easily understood from the
nature of idiotypic interactions: if stimulation by an external agent can
trigger a reorganization of the immune system based essentially on the same
stimulatory mechanisms, then a particularly aggressive strain of that agent can
find itself mimicked inside the immune system. The mimicking entities are the
so-called pathogenic idiotypes, which for about three decades (though often
under different denominations) have been recognized as an important cause of
autoimmune disease \cite{p83,o87,sm90}, sometimes in connection to
vaccine-related responses \cite{pmetal15}. The second key aspect on which we
concentrate is the interaction of the immune system with retroviruses
\cite{g07}, that is, RNA viruses that take advantage of a cell's internal medium
to turn RNA into a DNA precursor that eventually becomes part of the cell's own
DNA. Depending on the case, this can lead to the cell's destruction and to the
spread of the virus. This is the case of HIV when the cells in question are
cells of the immune system.

As it stands, nothing prevents the model we introduced in \cite{bds15} from
giving rise to autoimmunity-related phenomena. That model, however, has no
provisions for the explicit destruction of idiotypes by genotype action. Thus,
handling HIV-like retroviruses requires appropriate modifications to the model
(along with a new parameter, to be denoted by $\nu>0$). With these modifications
in place, the model can also be expected to give rise to a wider variety of
autoimmunity-related phenomena, since pathogenic idiotypes have been linked to
mutated DNA containing genes that become pathogenic in the manner of a
retrovirus but without any apparent connection to one. This is the case of the
so-called endogenous retroviruses \cite{s12,zengetal14}.

We proceed as follows. First we review our model of \cite{bds15} in
section~\ref{sec:model}, where we also introduce the $\nu$-dependent
modifications as well as new analytical results on a special case. Then we move
to a presentation of results in section~\ref{sec:results}, followed by
discussion in section~\ref{sec:disc} and conclusions in section~\ref{sec:concl}.

\section{Model}
\label{sec:model}

Each genotype or idiotype is represented by a length-$L$ sequence of $0$'s and
$1$'s. There are therefore $2^L$ distinct genotypes and $2^L$ distinct
idiotypes. The genotype consisting of only $0$'s is the wild type (the fittest
one, cf.\ section~\ref{sec:netdyn}).

\subsection{Network structure}

Our network has $2^{L+1}$ nodes, one for each of these entities. Its set of
edges is based on a directed random graph $D$ that depends on the probabilities
$p$ and $r$.

If nodes $i$ and $j$ are both genotypes, then an edge exists in $D$ directed
from $i$ to $j$ with probability $p_{ij}=p^{H_{ij}}$, where $H_{ij}$ is the
Hamming distance between the sequences defining $i$ and $j$ (i.e., the number of
loci at which they differ). The existence of this edge, in the case of
$i\neq j$, indicates that it is possible for $i$ to mutate into $j$ during
replication, so as expected, for fixed $p$ the connection probability between
two distinct genotypes grows as they become more similar. For $i=j$, the
mandatory self-loop at genotype $i$ indicates that it is possible for $i$ not to
mutate at all.

If nodes $i$ and $j$ are both idiotypes, then an edge from $i$ to $j$ exists in
$D$ with probability $r_{ij}=r^{L-H_{ij}}$, indicating when the edge does exist
that it is possible for $i$ to stimulate $j$ as part of the idiotypic dynamics
(this holds even if $i$ and $j$ are the same idiotype). Once again as expected
(now from the nature of the stimulation between idiotypes, based as it is on
molecular complementarity \cite{b07}), for fixed $r$ the connection probability
between two idiotypes grows as they become less similar. If $i$ and $j$ are
fully complementary (i.e., $H_{ij}=L$), then the edge is mandatory.

Graph $D$ contains further edges to account for the possibility of stimulation
of an idiotype by a genotype. For $i$ a genotype and $j$ an idiotype, an edge
exists from $i$ to $j$ with the same probability $r_{ij}$ as above, which is
fully justified by the fact that stimulation continues to be based on the exact
same complementarity principle.

\begin{figure}[t]
\begin{indented}
\item[]
\includegraphics[scale=1.20]{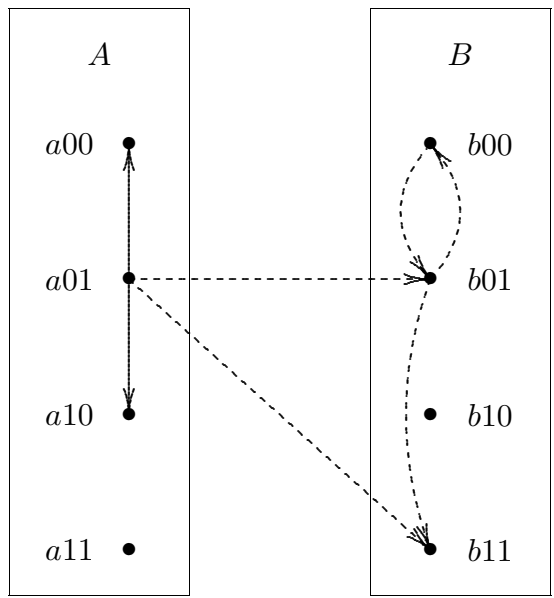}
\end{indented}
\caption{An instance of random graph $D$ for $L=2$, with genotype set
$A=\{a00,a01,a10,a11\}$ and idiotype set $B=\{b00,b01,b10,b11\}$. Genotype $a00$
is the wild type. Solid edges are related to genotype mutation by similarity;
dashed edges are related to idiotype stimulation by complementarity. This
instance has no self-loops on set $B$. The mandatory self-loops on set $A$ are
not shown, nor are the mandatory edges entailed by full complementarity inside
set $B$ or from set $A$ to set $B$. In-neighbor sets are
$I_{a00}=\{a00,a01\}$,
$I_{a01}=\{a01\}$,
$I_{a10}=\{a01,a10\}$,
$I_{a11}=\{a11\}$,
$I_{b00}=\{a11,b01,b11\}$,
$I_{b01}=\{a01,a10,b00,b10\}$,
$I_{b10}=\{a01,b01\}$, and
$I_{b11}=\{a00,a01,b00,b01\}$.
Out-neighbor sets are
$O_{a00}=\{a00,b11\}$,
$O_{a01}=\{a00,a01,a10,b01,b10,b11\}$,
$O_{a10}=\{a10,b01\}$,
$O_{a11}=\{a11,b00\}$,
$O_{b00}=\{b01,b11\}$,
$O_{b01}=\{b00,b10,b11\}$,
$O_{b10}=\{b01\}$, and
$O_{b11}=\{b00\}$.}
\label{fig1}
\end{figure}

Given a fixed instance of graph $D$ (that is, a deterministic realization of
$D$ that may contain some of the nonmandatory edges but not others), we use
$I_i$ to denote the set of in-neighbors of node $i$ and $O_i$ to denote its set
of out-neighbors. We partition the graph's set of $2^{L+1}$ nodes into a set $A$
containing the $2^L$ genotypes and a set $B$ containing the $2^L$ idiotypes.
Note that, by the definition of $D$, $I_i$ has a nonempty intersection with $A$
but not with $B$ if $i\in A$. Similarly, given $i\in B$ it follows that $O_i$
has a nonempty intersection with $B$ but not with $A$. The sets $O_i\cap A$ and
$O_i\cap B$ for $i\in A$, and also $I_i\cap A$ and $I_i\cap B$ for $i\in B$, are
all necessarily nonempty. We exemplify a $D$ instance in figure~\ref{fig1} (here
reproduced from \cite{bds15} for the reader's benefit).

\subsection{Network dynamics}
\label{sec:netdyn}

Given an instance of random graph $D$, the network dynamics is described by a
set of coupled differential equations, one for each of the nodes, each giving
the rate at which the corresponding genotype or idiotype's abundance varies with
time. A first form of these equations refers to absolute abundances, $X_i$ for
genotype or idiotype $i$.

For $i\in A$ (i.e., $i$ is a genotype), the rate at which $X_i$ grows depends on
the genotypes' fitnesses. We assume the fitness of genotype $j$ to decay
exponentially from that of the wild type, assumed to be $1$, as a function of
the number of loci at which the two genotypes differ. Denoting the fitness of
genotype $j$ by $f_j$, we have $f_j=2^{-d_j}$, where $d_j$ is the number of
$1$'s in the sequence representing genotype $j$. The growth rate of $X_i$ also
depends on the probability that genotype $j\in I_i$ mutates into genotype $i$,
denoted by $q_{ji}$ and assumed proportional to $p_{ji}$ in such a way that
$\sum_{k\in O_j\cap A}q_{jk}=1$, and on the probability that genotype $i$
stimulates idiotype $j\in O_i\cap B$, denoted by $s_{ij}$ and assumed
proportional to $r_{ij}$ in such a way that $\sum_{k\in O_i\cap B}s_{ik}=1$.
Given the rate $\mu>0$ at which genotype abundances get reduced by the action of
the idiotypes, we have
\begin{equation}
\dot{X}_i=\sum_{j\in I_i}f_jq_{ji}X_j-\mu\sum_{j\in O_i\cap B}s_{ij}X_j.
\label{eq:XA}
\end{equation}
Had the second summation been absent, this would be the well-known quasispecies
equation \cite{be06}, written for the $D$ instance at hand.

For $i\in B$ (i.e., $i$ is an idiotype), the growth rate of $X_i$ depends on the
same stimulation probability $s_{ji}$ as above, where $j$ is either an idiotype
or a genotype. Denoting by $\lambda>0$ the rate at which idiotypes proliferate
in response to stimulation by genotypes or idiotypes, and by $\nu$ the rate at
which idiotype abundances get reduced due to the stimulation by genotypes, we
have
\begin{eqnarray}
\dot{X}_i
&=&\lambda\sum_{j\in I_i}s_{ji}X_j-\nu\sum_{j\in I_i\cap A}s_{ji}X_j\cr
&=&(\lambda-\nu)\sum_{j\in I_i\cap A}s_{ji}X_j
+\lambda\sum_{j\in I_i\cap B}s_{ji}X_j.
\label{eq:XB}
\end{eqnarray}
For $\nu=0$, this is the equation that in \cite{bds15} governs the growth of
idiotype $i$'s absolute abundances. For $\nu=\lambda$, the evolution of $X_i$
gets decoupled from the influence of any genotype.

A more useful form of equations~(\ref{eq:XA}) and~(\ref{eq:XB}) can be obtained
by considering relative, rather than absolute, abundances. Rewriting the two
equations in these terms leads to the appearance of further terms that reflect
the removal of genotypes as they mutate into other genotypes and of idiotypes as
they stimulate (and consequently get recognized and then destroyed by) one
another. For $x_i$ the relative abundance of genotype or idiotype $i$, we
achieve this by letting
$x_i=X_i/\sum_{k\in A\cup B}X_k$ for $i\in A\cup B$, whence
$\sum_{i\in A\cup B}x_i=1$. Denoting by $x_A$ the total relative abundance of
genotypes,
\begin{equation}
x_A=\sum_{i\in A}x_i,
\end{equation}
yields
\begin{eqnarray}
\dot{x}_i
&=&\frac{\dot{X}_i}{\sum_{k\in A\cup B}X_k}
-x_i\frac{\sum_{k\in A\cup B}\dot{X}_k}{\sum_{k\in A\cup B}X_k}\cr
&=&\frac{\dot{X}_i}{\sum_{k\in A\cup B}X_k}
-x_i(\phi-\mu\psi+\lambda-\nu x_A),
\label{eq:x*}
\end{eqnarray}
where $\phi=\sum_{k\in A}f_kx_k$ and
$\psi=\sum_{k\in B}x_k\sum_{\ell\in I_k\cap A}s_{\ell k}$.

The equations for relative abundances are then
\begin{equation}
\dot{x}_i
=\sum_{j\in I_i}f_jq_{ji}x_j
-\mu\sum_{j\in O_i\cap B}s_{ij}x_j
-x_i(\phi-\mu\psi+\lambda-\nu x_A)
\label{eq:xA}
\end{equation}
for $i\in A$ and
\begin{equation}
\dot{x}_i
=(\lambda-\nu)\sum_{j\in I_i\cap A}s_{ji}x_j
+\lambda\sum_{j\in I_i\cap B}s_{ji}x_j
-x_i(\phi-\mu\psi+\lambda-\nu x_A)
\label{eq:xB}
\end{equation}
for $i\in B$. Note that
$\phi/\sum_{k\in A}x_k$ is the average genotype fitness. In a similar vein, in
\cite{bds15} we refer to $\psi/\sum_{k\in B}x_k$ as the average idiotype
proliferability. Setting $\nu=\lambda$ in equation~(\ref{eq:xB}) decouples
idiotype evolution from direct genotype action, similarly to what happens in
equation~(\ref{eq:XB}). In fact, letting $x_B$ be the total relative abundance
of idiotypes,
\begin{equation}
x_B=\sum_{i\in B}x_i=1-x_A,
\end{equation}
allows equation~(\ref{eq:xB}) to be rewritten as
\begin{equation}
\dot{x}_i
=\lambda\sum_{j\in I_i\cap B}s_{ji}x_j-x_i(\phi-\mu\psi+\lambda x_B)
\end{equation}
for $\nu=\lambda$, where it also becomes clear that the decoupling can never be
complete, since the idiotype population is continuously influenced by that of
genotypes through the renormalizing effect of $\phi$.

\subsection{A special case}\label{sec:model-special}

Equations~(\ref{eq:xA}) and~(\ref{eq:xB}) are in general intractable
analytically, but assuming the deterministic variant of graph $D$ in which all
possible edges really do exist, and moreover that all genotype fitnesses are the
same as the wild type's, leads to a special case that is amenable to analytical
solution for $x_A$ (hence for $x_B$). These changes require that we set $p=1$
and $r=1$, which leads to $q_{ij}=r_{ij}=1/2^L$ for all meaningful pairings of
$i$ and $j$ and also to $\psi=x_B$, and moreover that we adopt $f_j=1$ for all
$j\in A$, which leads to $\phi=x_A$. Combined, these simplifying assumptions
allow equation~(\ref{eq:xA}) to be rewritten as
\begin{equation}
\dot{x}_i
=(1/2^L)x_A-(\mu/2^L)x_B-x_i[(1-\nu)x_A-\mu x_B+\lambda],
\label{eq:xAs}
\end{equation}
where $i\in A$, and equation~(\ref{eq:xB}) as
\begin{equation}
\dot{x}_i
=[(\lambda-\nu)/2^L]x_A+(\lambda/2^L)x_B -x_i[(1-\nu)x_A-\mu x_B+\lambda],
\label{eq:xBs}
\end{equation}
where $i\in B$.

Summing up equation~(\ref{eq:xAs}) on $i\in A$ and equation~(\ref{eq:xBs}) on
$i\in B$ yields, respectively,
\begin{equation}
\dot{x}_A
=(1-\lambda)x_A-\mu x_B-[(1-\nu)x_A-\mu x_B]x_A
\label{eq:xAt}
\end{equation}
and
\begin{equation}
\dot{x}_B
=(\lambda-\nu)x_A-[(1-\nu)x_A-\mu x_B]x_B.
\end{equation}
Using the fact that by definition $x_A+x_B=1$ at all times, it is a simple
matter to check that, as expected, $\dot{x}_A+\dot{x}_B=0$ also at all times.

For $\alpha=(1+2\mu-\lambda)/\gamma$ and $\beta=\mu/\gamma$, with
$\gamma=1+\mu-\nu$, we can rewrite equation~(\ref{eq:xAt}) as
\begin{equation}
\dot{x}_A=-\gamma(x_A-x_A^+)(x_A-x_A^-).
\label{eq:xAt2}
\end{equation}
In this equation, $x_A^+$ and $x_A^-$ are the two roots of
$x_A^2-\alpha x_A+\beta=0$, that is,
\begin{eqnarray}
x_A^+,x_A^-
&=&\frac{\alpha\pm\sqrt{\alpha^2-4\beta}}{2}\cr
&=&\frac{1+2\mu-\lambda\pm\sqrt{(1-\lambda)^2+4\mu(\nu-\lambda)}}{2(1+\mu-\nu)}.
\label{eq:roots}
\end{eqnarray}

These roots are finite real numbers if $\gamma\neq 0$ and
$\alpha^2-4\beta\ge 0$, that is, if $\nu\neq 1+\mu$ and
$(1-\lambda)^2+4\mu(\nu-\lambda)\ge 0$. This requires either
$\lambda\le 1+2\mu-2\sqrt{\mu\gamma}$ or $\lambda\ge 1+2\mu+2\sqrt{\mu\gamma}$,
provided $\gamma>0$ in either case, that is, provided $\nu<1+\mu$. The latter
also implies that $\beta>0$, in which case both roots are nonzero and have the
same sign. They are moreover positive if $\alpha>0$, that is, if
$\lambda<1+2\mu$. If in addition $\nu\le\lambda$, then $x_A^+\le 1$.

Equation~(\ref{eq:xAt2}) can be used to obtain the value of $x_A$ in the limit
as $t\to\infty$. This depends on how $x_A(0)$, the value of $x_A$ at $t=0$,
relates to $x_A^-$, as can be seen by considering the following three cases.
The first one is that in which $x_A(0)<x_A^-$, implying $\dot{x}_A<0$, hence
$x_A=0$ in the limit. The second case is that of $x_A(0)=x_A^-$, which clearly
yields $x_A=x_A^-$ at all times. The third case, finally, is that of
$x_A(0)>x_A^-$, and then $\dot{x}_A$ is constrained by how $x_A$ relates to
$x_A^+$, yielding in all cases $x_A=x_A^+$ in the limit.

It is important to note that our conclusion of $x_A=0$ as the limiting value
whenever $x_A(0)<x_A^-$ is a consequence of the implicit constraint that
$x_A\ge 0$ at all times. Likewise, setting $\nu>\lambda$ while ensuring that
both $x_A^-$ and $x_A^+$ are positive, finite real numbers leads to $x_A^+>1$,
so the conclusion that $x_A=x_A^+$ in the limit for $x_A(0)>x_A^-$ is subject to
the further implicit constraint that $x_A\le 1$ at all times and should in this
case be read as $x_A=1$. These constraints are nowhere accounted for during the
formulation of the special case, only when bounding possible parameter values,
but clearly they become manifest when we consider the effect of $x_A(0)$ on
limiting values of $x_A$. A similar issue is raised in
section~\ref{sec:results}.

In section~\ref{sec:disc}, we return to this special case of a fully connected
network (i.e., $p=r=1$) and equally fit genotypes, and show that the analytical
results obtained for this case can sometimes be used as quite reasonable
approximations in more plausible scenarios. This is remarkable, since in general
we have $p,r\ll 1$ as well as genotype fitnesses that decay exponentially from
that of the wild type.

\section{Results}
\label{sec:results}

We begin by recognizing, as in \cite{bds15} for the case of $\nu=0$, that
$\dot{X}_i$ can be negative when $X_i=0$ in both equations~(\ref{eq:XA})
and~(\ref{eq:XB}), thus violating the implicit constraint that $X_i\ge 0$ at all
times. We prevent this by rewriting those equations as
\begin{equation}
\dot{X}_i=\sum_{j\in I_i}f_jq_{ji}X_j-\mu H(X_i)\sum_{j\in O_i\cap B}s_{ij}X_j
\end{equation}
and
\begin{equation}
\dot{X}_i
=[\lambda-\nu H(X_i)]\sum_{j\in I_i\cap A}s_{ji}X_j
+\lambda\sum_{j\in I_i\cap B}s_{ji}X_j,
\end{equation}
respectively, where $H(z)$ is the Heaviside step function, slightly modified to
yield $1$ if $z>0$ and $0$ otherwise.

The consequence of this for equations~(\ref{eq:xA}) and~(\ref{eq:xB}) is that
they get reformulated as well, becoming
\begin{equation}
\dot{x}_i
=\sum_{j\in I_i}f_jq_{ji}x_j
-\mu H(x_i-\delta)\sum_{j\in O_i\cap B}s_{ij}x_j
-x_i(\phi-\mu\psi+\lambda-\nu\xi)
\label{eq:impl1}
\end{equation}
and
\begin{eqnarray}
\dot{x}_i
=[\lambda-\nu H(x_i-\delta)]\sum_{j\in I_i\cap A}s_{ji}x_j
&+&\lambda\sum_{j\in I_i\cap B}s_{ji}x_j\nonumber\\
&-&x_i(\phi-\mu\psi+\lambda-\nu\xi),
\label{eq:impl2}
\end{eqnarray}
respectively, where $\psi$ is now written as
\begin{equation}
\psi=\sum_{k\in B}x_k\sum_{\ell\in I_k\cap A}s_{\ell k}H(x_\ell-\delta)
\label{eq:impl3}
\end{equation}
and
\begin{equation}
\xi=\sum_{k\in A}x_k\sum_{\ell\in O_k\cap B}s_{k\ell}H(x_\ell-\delta).
\label{eq:impl4}
\end{equation}
The $\delta$ appearing in equations~(\ref{eq:impl1}) through~(\ref{eq:impl4}) is
a small positive constant (we use $\delta=10^{-10}$) meant to prevent
instabilities during numerical integration. It also affects the determination of
the time step to be used in each iteration, as detailed in \cite{bds15}.

\begin{figure}[t]
\begin{indented}
\item[]
\includegraphics[scale=0.45]{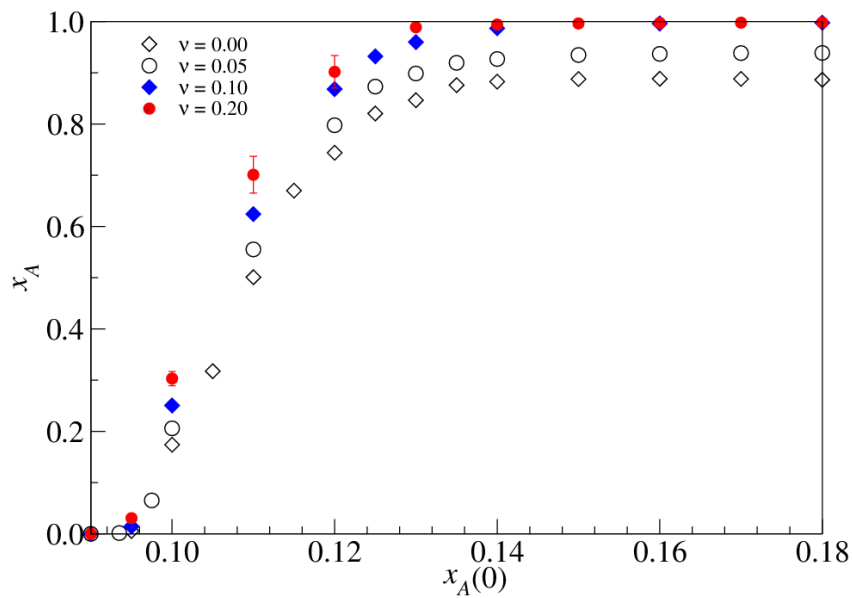}
\end{indented}
\caption{Stationary-state relative abundance of genotypes ($x_A$) as a function
of its initial value ($x_A(0)$). Data are given for $p=r=0.1$ and
$\lambda=\mu=0.1$.}
\label{fig2}
\end{figure}

Our results are given for $L=10$ (hence $1\,024$ genotypes and $1\,024$
idiotypes) and appear in figures~\ref{fig2}--\ref{fig9}. In these figures we
continue to explore the parameter niche that in \cite{bds15} was found to yield
informative results, now enriched by various possibilities for the new
parameter, $\nu$. Most results refer explicitly to the total relative abundance
of genotypes, $x_A$, either highlighting the evolution in time of the associated
probability density (figures~\ref{fig3}--\ref{fig5}) or the behavior of the
corresponding stationary-state expected value with respect to some parameter
(all other figures, except figure~\ref{fig8}). The results in figure~\ref{fig8}
constitute the only exception and refer explicitly to the behavior of $x_B$, the
total relative abundance of idiotypes, aiming to highlight the appearance of
pathogenic idiotypes.

As in section~\ref{sec:model-special}, we use $x_A(0)$ to denote the initial
value of $x_A$. All results come from time-stepping equations~(\ref{eq:impl1})
and~(\ref{eq:impl2}) through $t=20$. This upper bound on $t$, though
substantially lower than the one in \cite{bds15}, was empirically found to allow
a nearly stationary state to be reached in all cases. This lowering has been
instrumental in allowing all our results to be obtained within a reasonable
amount of time, since it has turned out that solving the equations for
$\nu>0$ is substantially more time-consuming than for the $\nu=0$ scenarios of
\cite{bds15}. Such additional demand for computation time has also resulted in
the adoption of substantially fewer $D$ instances per parameter configuration in
comparison to \cite{bds15}. Error bars are then used in the figures whenever
legible. Initial conditions were $x_i=x_A(0)/2^L$ for $i\in A$ and
$x_i=[1-x_A(0)]/2^L$ for $i\in B$.

The vast majority of our results come from using $x_A(0)=0.1$, along with a
base set of parameter values that we perturb to obtain different scenarios. This
base set consists of $p=r=0.1$, $\lambda=\mu=0.1$, and $\nu=0.1$. Our choice of
$x_A(0)=0.1$ for most cases comes from examining figure~\ref{fig2}, where the
stationary-state value of $x_A$ is plotted against $x_A(0)$ for the base set of
parameters enlarged by three further values of $\nu$
($0.00$, $0.05$, and $0.2$, the first of these providing a connection with the
work in \cite{bds15}). It is clear from the figure that $x_A(0)=0.1$ is located
right past a transition from the expected destruction to the expected survival
of the quasispecies, therefore well positioned for a wide variety of behaviors
to ensue.

\section{Discussion}
\label{sec:disc}

In addition to its use in assisting with the choice of $x_A(0)$,
figure~\ref{fig2} highlights one of the main roles played by the new parameter,
$\nu$, namely to provide ever increasing chances of quasispecies survival as it
is itself increased. Not only this, but any combination of sufficiently high
$x_A(0)$ and $\nu$ values seem to imply the complete shattering of the idiotypic
network, since for such combinations the stationary-state value of $x_A$ tends
to $1$ and that of $x_B$ to $0$. However, assigning nonzero values to $\nu$ has
additional, more subtle consequences. We explore them in what follows.

\begin{figure}[p]
\begin{indented}
\item[]
\includegraphics[scale=0.45]{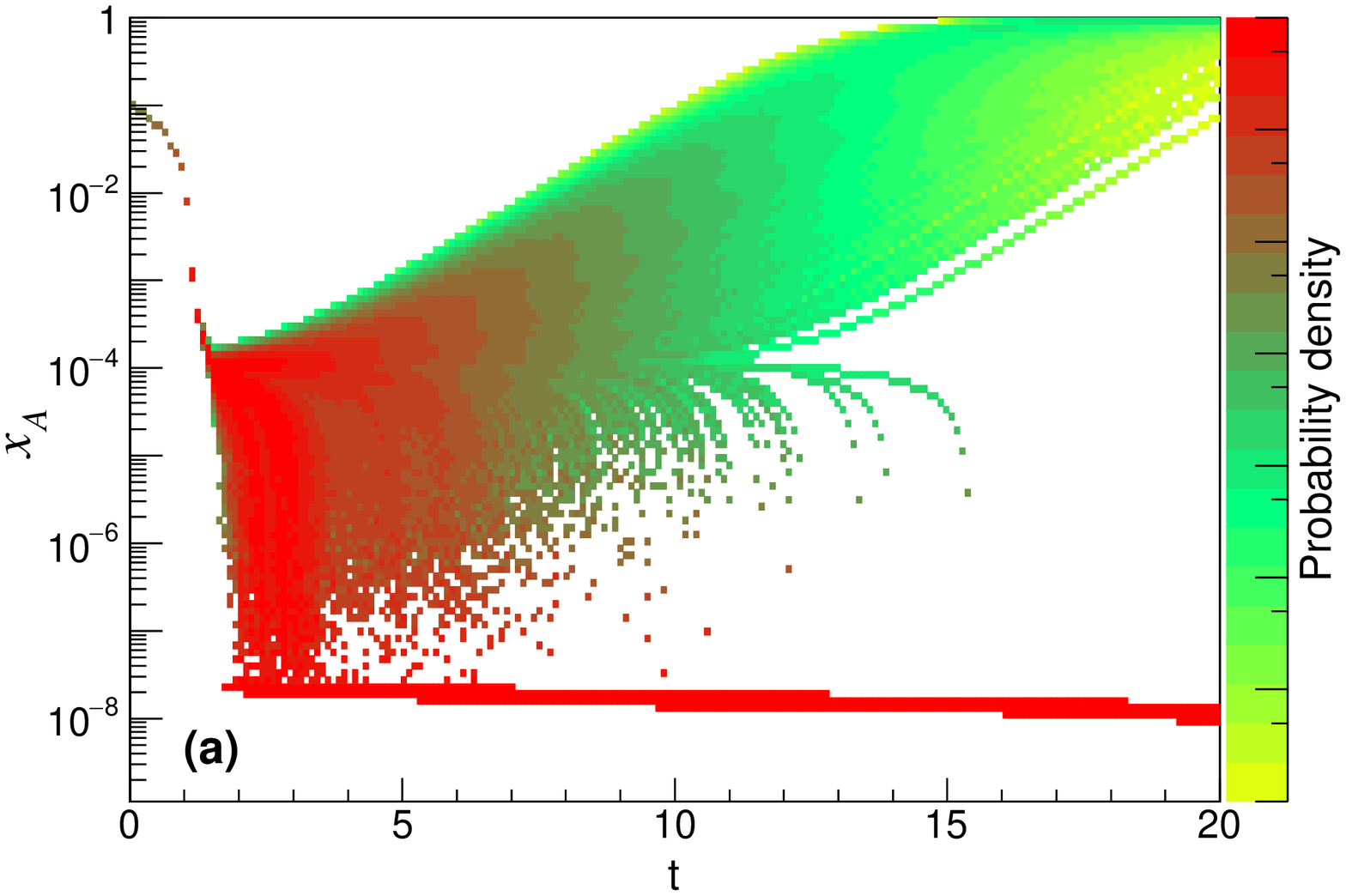}\\
\includegraphics[scale=0.45]{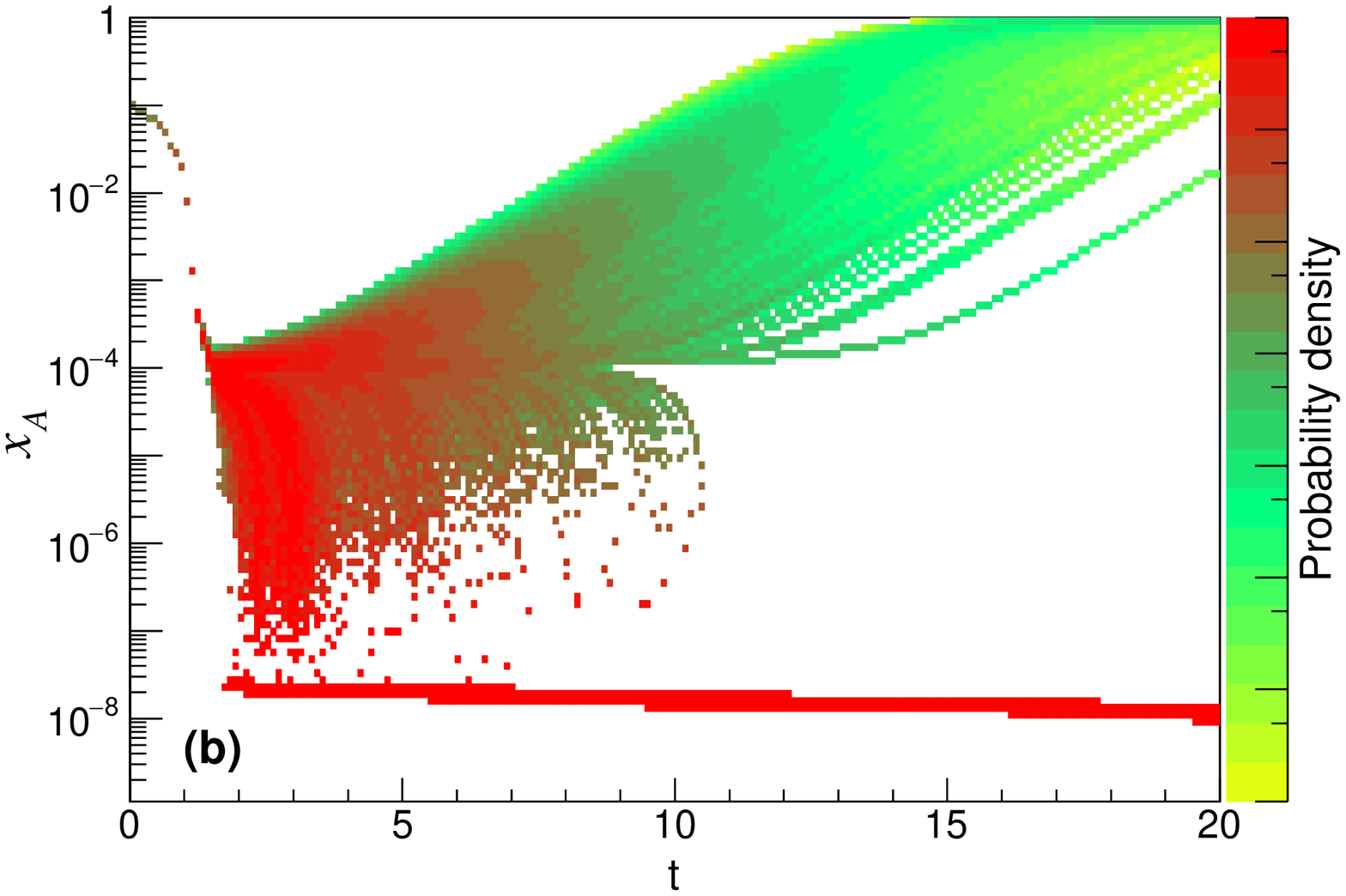}\\
\includegraphics[scale=0.45]{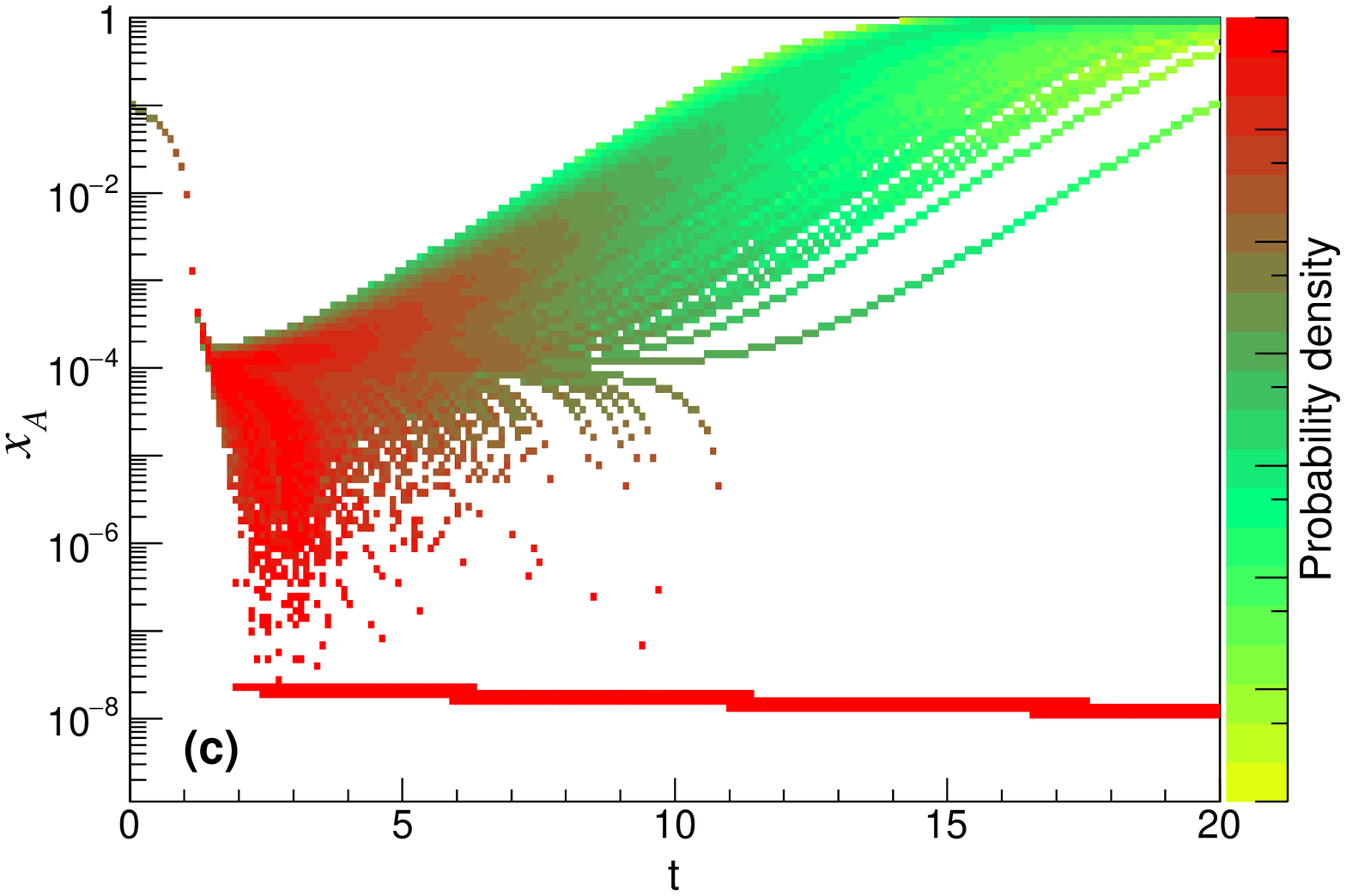}
\end{indented}
\caption{Evolution of the probability density of $x_A$ from $x_A(0)=0.1$, for
$\nu=0.05$ (a), $\nu=0.1$ (b), and $\nu=0.2$ (c), with $p=r=0.1$ and
$\lambda=\mu=0.1$. Data are log-binned to the base $1.2$. Probability densities
are given according to the color-coded logarithmic scale on the right of each
panel, ranging from $10^{-3}$ (at the bottom of the scale) to $10^4$ (at the
top).}
\label{fig3}
\end{figure}

\begin{figure}[p]
\begin{indented}
\item[]
\includegraphics[scale=0.45]{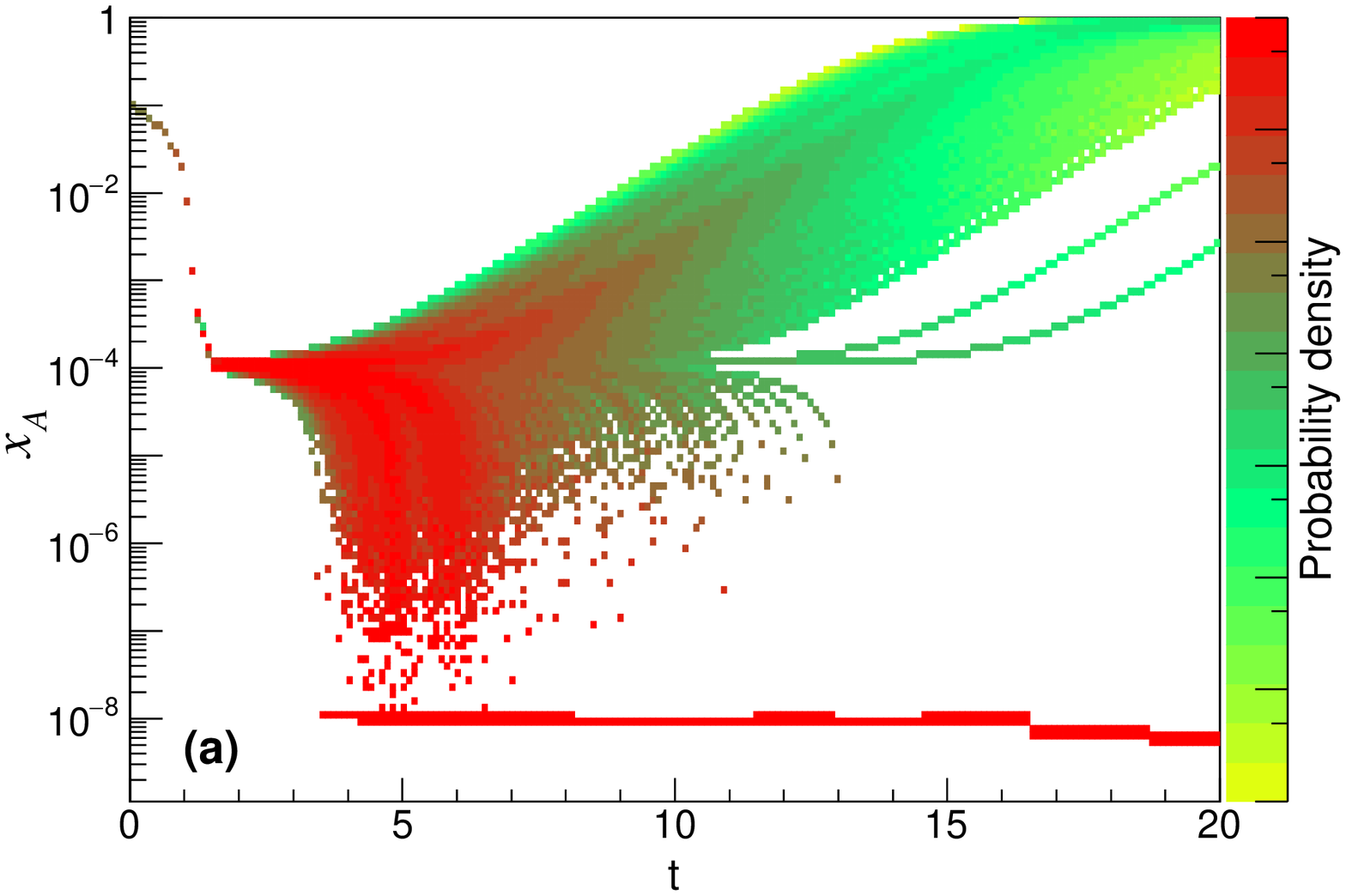}\\
\includegraphics[scale=0.45]{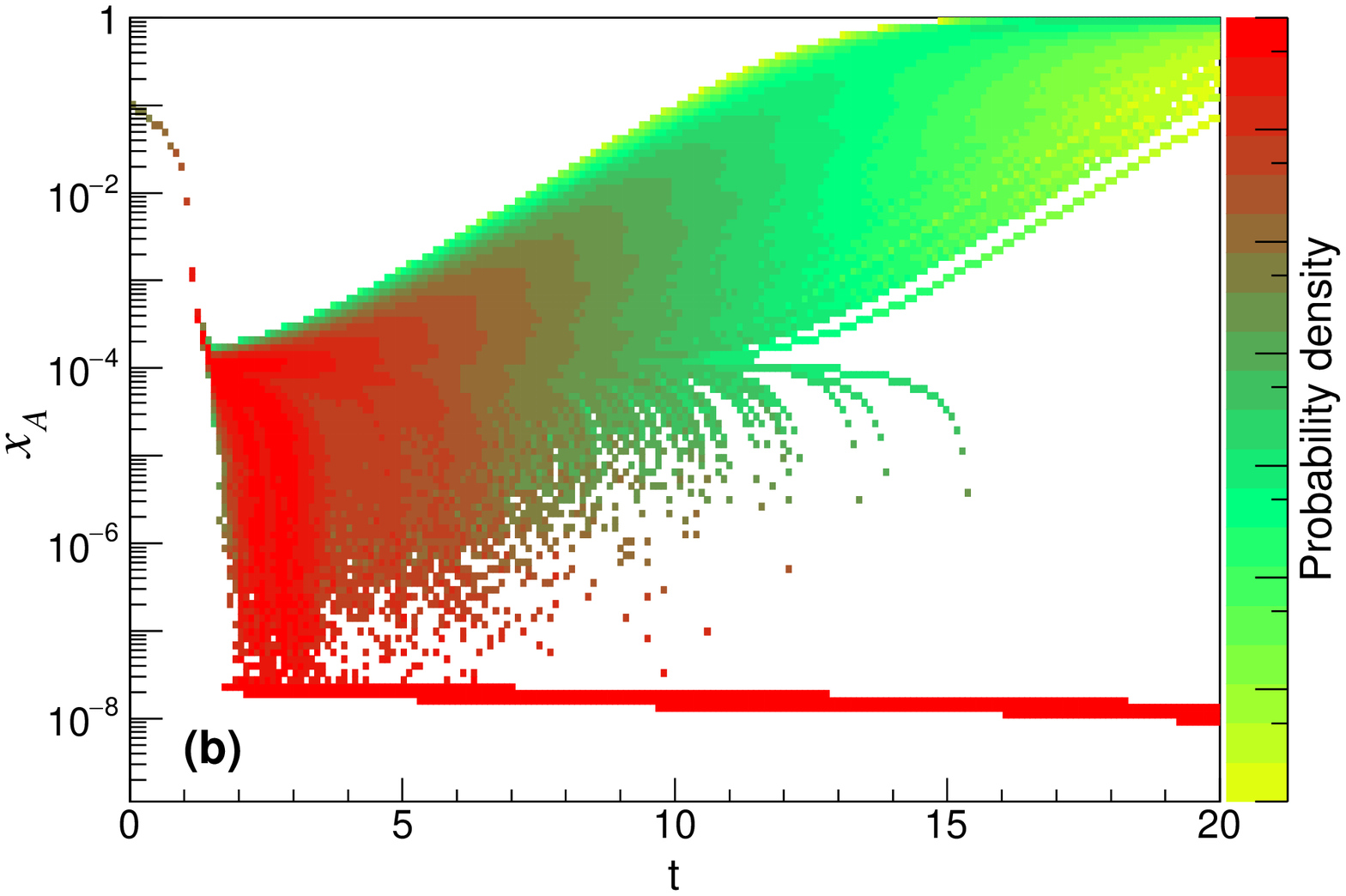}\\
\includegraphics[scale=0.45]{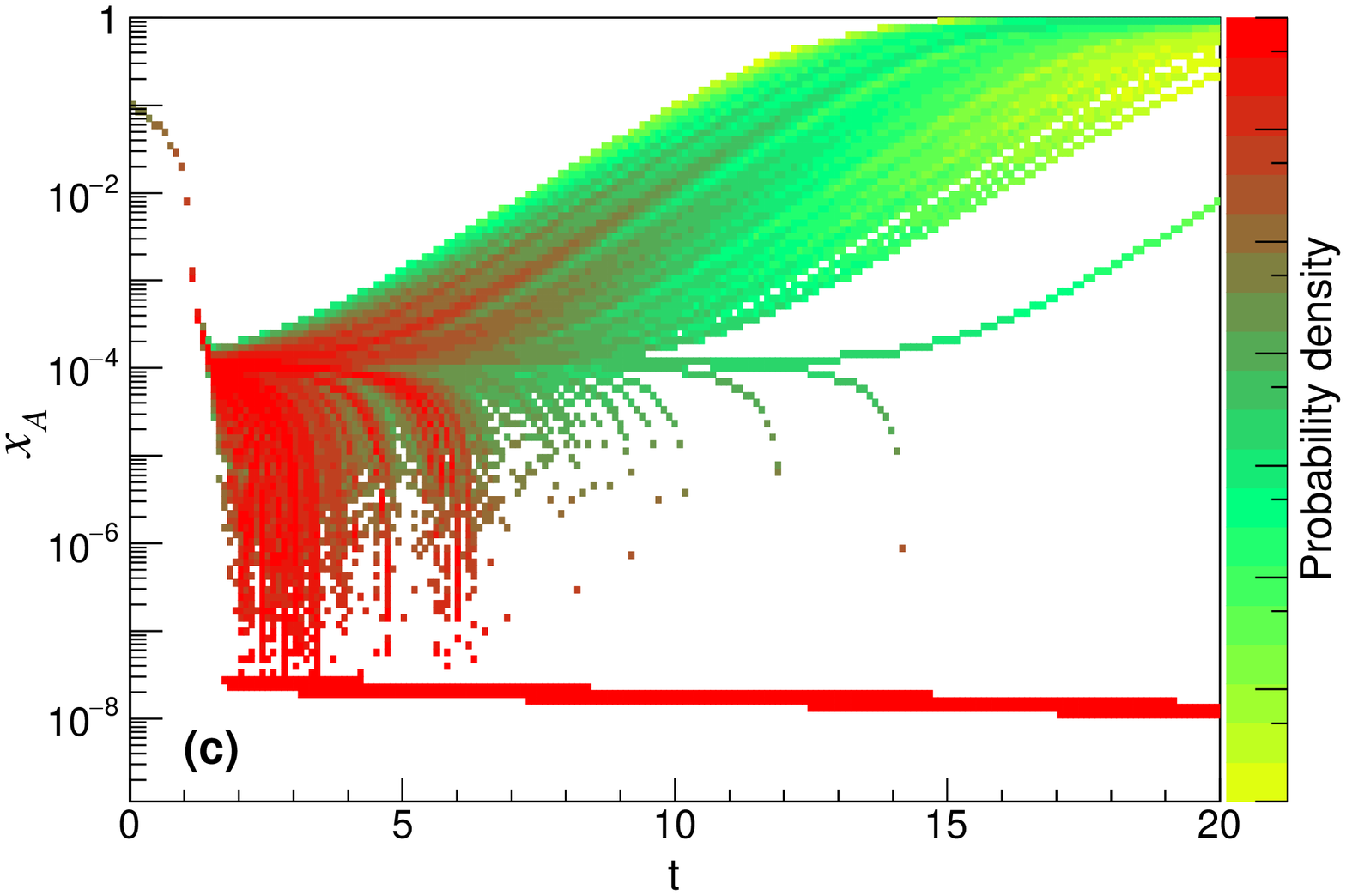}
\end{indented}
\caption{Evolution of the probability density of $x_A$ from $x_A(0)=0.1$, for
$p=0.01$, $r=0.1$ (a), $p=r=0.1$ (b), and $p=0.1$, $r=0.01$ (c), with 
$\lambda=\mu=0.1$ and $\nu=0.05$. Data binning and color codes are as in
figure~\ref{fig3}.}
\label{fig4}
\end{figure}

\begin{figure}[p]
\begin{indented}
\item[]
\includegraphics[scale=0.45]{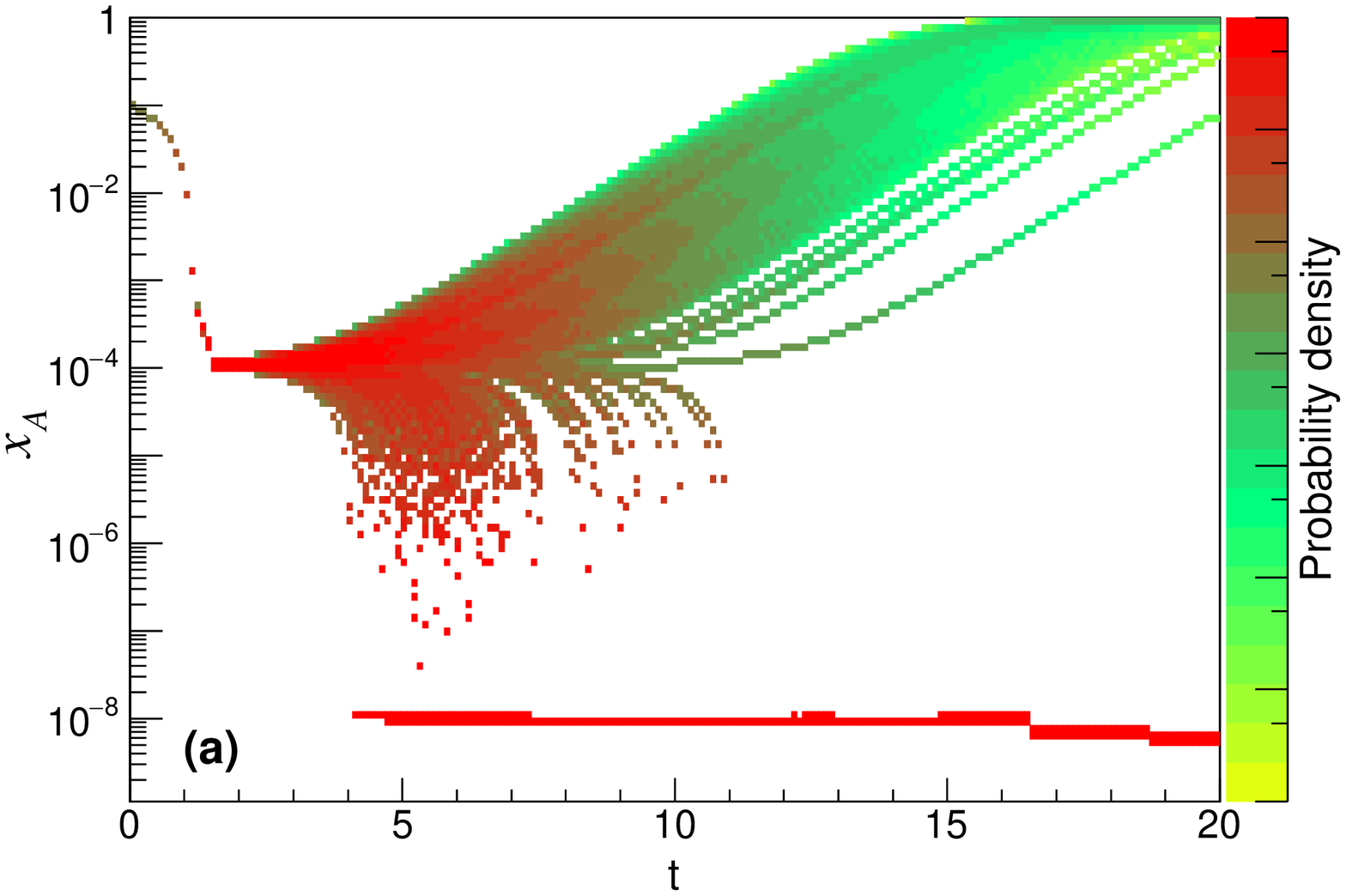}\\
\includegraphics[scale=0.45]{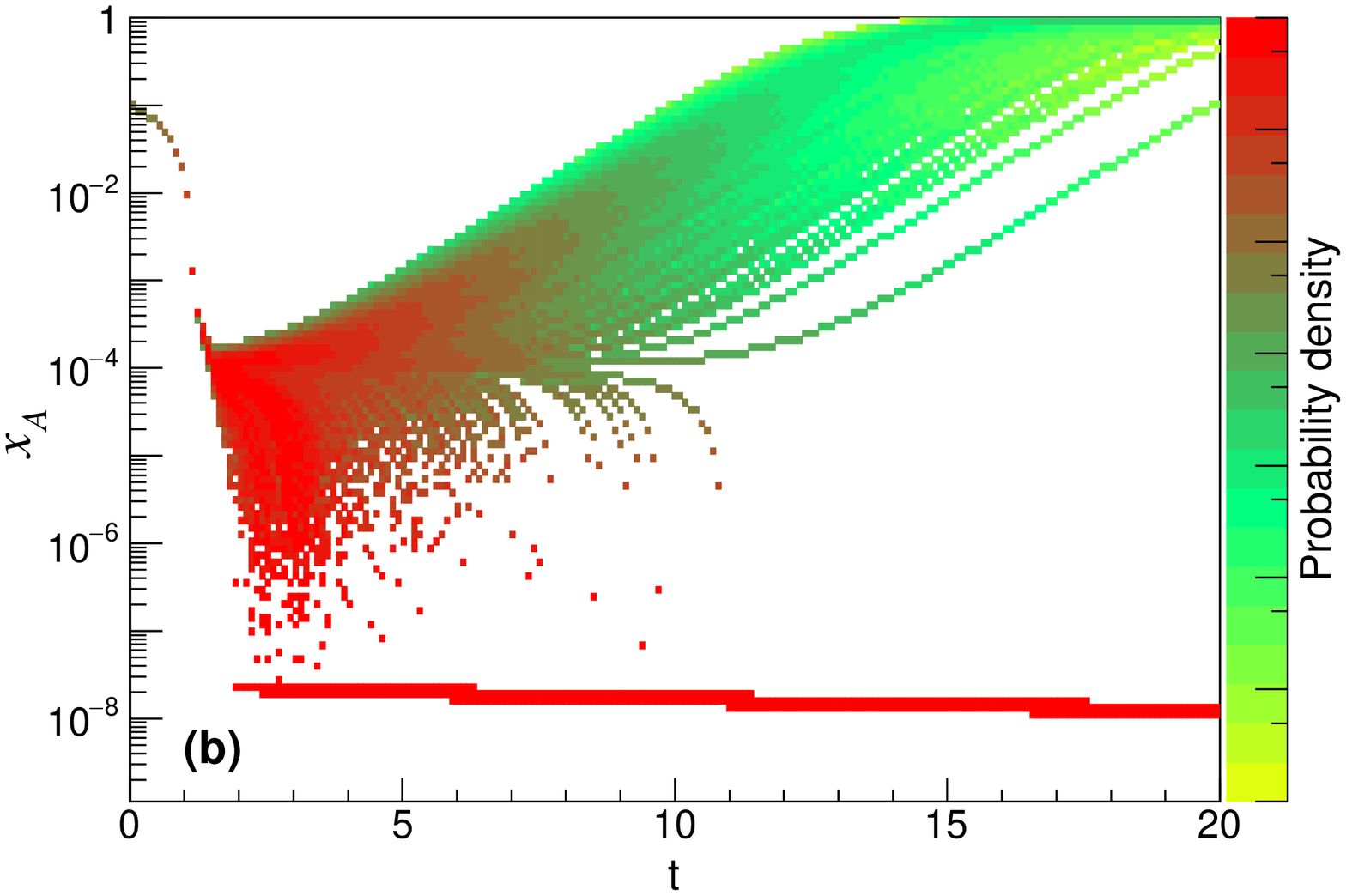}\\
\includegraphics[scale=0.45]{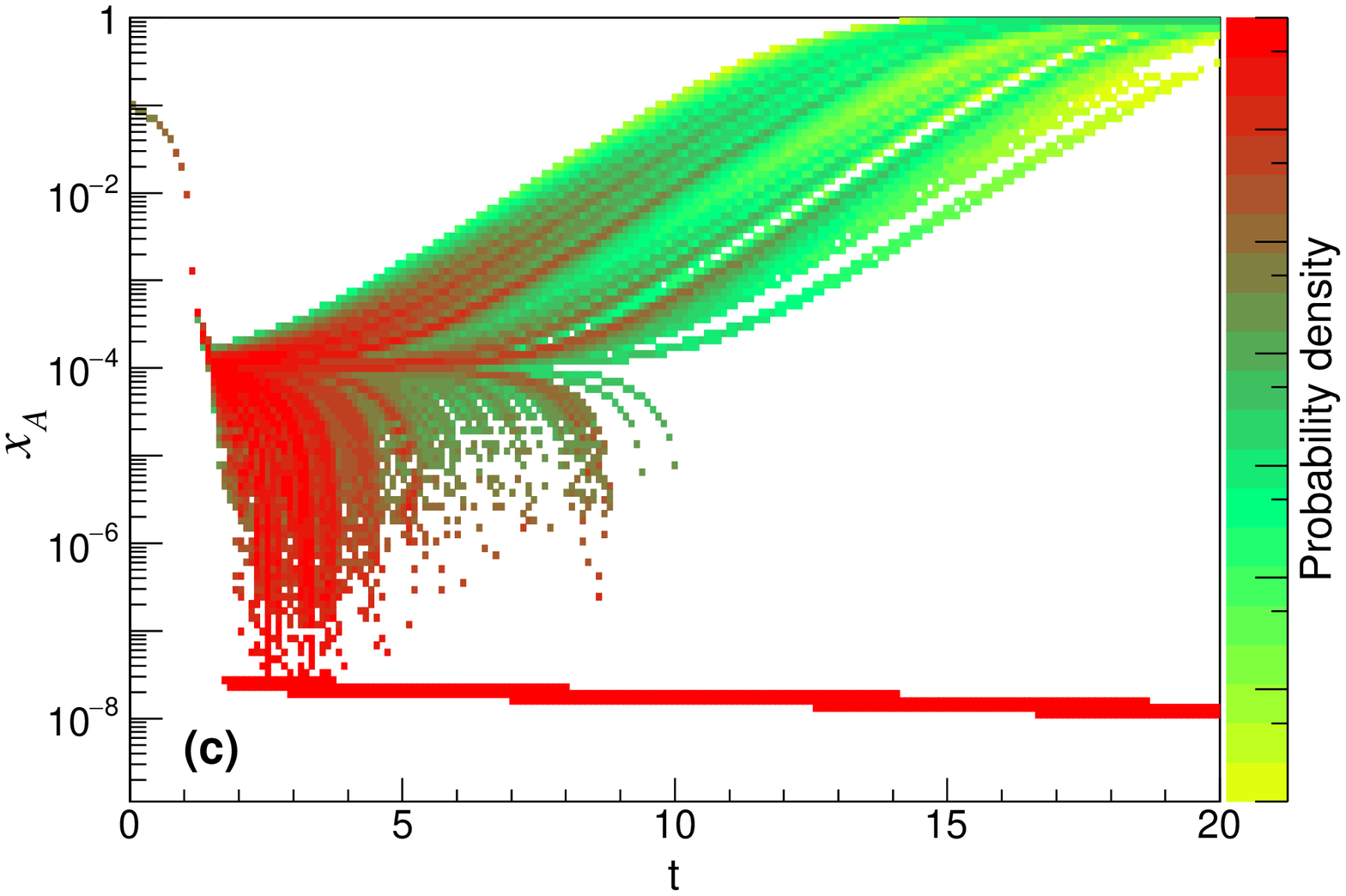}
\end{indented}
\caption{Evolution of the probability density of $x_A$ from $x_A(0)=0.1$, for
$p=0.01$, $r=0.1$ (a), $p=r=0.1$ (b), and $p=0.1$, $r=0.01$ (c), with 
$\lambda=\mu=0.1$ and $\nu=0.2$. Data binning and color codes are as in
figure~\ref{fig3}.}
\label{fig5}
\end{figure}

All panels in figures~\ref{fig3}--\ref{fig5} show the probability density of
$x_A$ as $t$ is varied from $t=0$ through $t=20$, in all cases having
$x_A(0)=0.1$ and $\lambda=\mu=0.1$, the latter meaning that the rate of idiotype
proliferation due to stimulation by both genotypes and idiotypes is the same as
the rate of genotype removal as genotypes stimulate idiotypes. In other words,
with $\lambda=\mu$ the immune system is as responsive in reorganizing itself
when external stimulation by genotypes occurs as it is in removing those very
genotypes. This seems like an ideal setting in which to investigate the effect
of the new parameter, $\nu$, since it is through this parameter that genotypes
acquire the ability of destroying, in addition to stimulating, idiotypes. Each
of the three figures differs from the other two in how the remaining parameters
($p$, $r$, and $\nu$) are valued.

One trait that is common to all panels in these three figures, parameter values
not withstanding, is that the probability density of $x_A$ does not depend on
the particular instance of graph $D$ in question up to about $t=2$. That is, up
to this time the density is sharply concentrated and reflects a strong decline
in the value of $x_A$ relative to its initial value. Thereafter the structure of
graph $D$ begins to exert its influence and a much greater variety of behaviors
is observed, including in all cases a fraction of $D$ instances for which the
quasispecies does not survive. 

Figure~\ref{fig3} has $p=r=0.1$, meaning that on average a genotype can mutate
into roughly as many genotypes as there are idiotypes that it can stimulate,
this number being practically the same as the number of other idiotypes that on
average an idiotype can stimulate (cf.\ \cite{bds15}, section 2.6). It also
means, in terms of the dynamics governing how the genotype and idiotype
populations evolve, that the same underlying probability is used. The value of
$\nu$ varies from $\nu=0.05$ in panel (a), to $\nu=0.1$ in panel (b), to
$\nu=0.2$ in panel (c), each new value causing the ratio $\nu/\lambda$ to double
(from $0.5$, to $1$, to $2$). This ratio indicates how greater the rate of
idiotype removal by genotypes is than the rate at which idiotypes proliferate by
virtue of being stimulated. As noted in section~\ref{sec:netdyn}, the middle
ground represented by $\nu=\lambda$ in panel (b) entails a regime of idiotype
evolution that does not directly depend on the genotype side of the network
(though, conversely, idiotypes continue to drive the destruction of genotypes
through the $\mu$ parameter).

\begin{figure}[t]
\begin{indented}
\item[]
\includegraphics[scale=0.45]{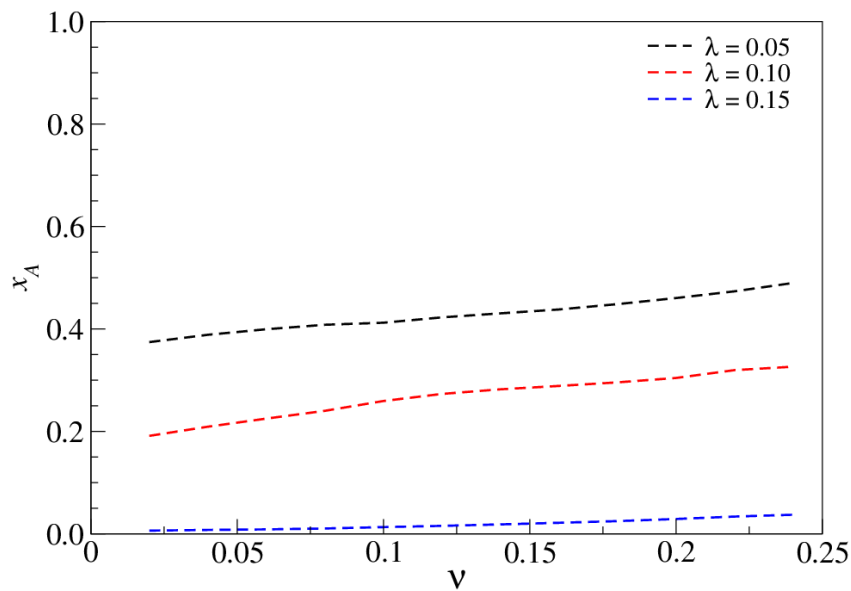}
\end{indented}
\caption{Stationary-state relative abundance of genotypes ($x_A$) as a function
of $\nu$. Data are given for $p=r=0.1$, $\mu=0.1$, and $x_A(0)=0.1$.}
\label{fig6}
\end{figure}

Figure~\ref{fig3} seems to indicate that increasing the ratio $\nu/\lambda$
facilitates the survival of the quasispecies by concentrating more probability
mass at the higher values of $x_A$. To see that this is indeed so, we resort to
the additional data shown in figure~\ref{fig6}, where the stationary-state
average of $x_A$ over all $D$ instances is shown as a function of $\nu$.
Figure~\ref{fig6} covers all parameter values used in figure~\ref{fig3}, as well
as two additional values of $\lambda$ ($0.05$ and $0.15$) and several values of
$\nu$ in addition to the three values to which figures~\ref{fig3}(a--c)
correspond. The new data confirm the conclusion above, namely, that increasing
the ratio $\nu/\lambda$ is beneficial to the quasispecies. This can be seen both
by fixing $\lambda$ (fixing one of the three plots in the figure~\ref{fig6})
while $\nu$ is increased and by fixing $\nu$ while decreasing $\lambda$ (moving
from the bottommost plot to the topmost).

Both figure~\ref{fig4} and figure~\ref{fig5} have $\lambda=\mu=0.1$, as in
figure~\ref{fig3}, but differ from that figure in that the $p/r$ ratio increases
by a factor of $10$ from panel (a) ($p=0.01$, $r=0.1$) to panel (b) ($p=r=0.1$),
and once again by the same factor from panel (b) to panel (c) ($p=0.1$,
$r=0.01$). They differ from each other in that $\nu=0.05$ in figure~\ref{fig4}
and $\nu=0.2$ in figure~\ref{fig5}. For comparison's sake, therefore, they are
both anchored in figure~\ref{fig3}, whose panels (a) and (c) are identical to
figures~\ref{fig4}(b) and~\ref{fig5}(b), respectively.

Increasing the $p/r$ ratio has consequences for the structure of graph $D$.
Specifically, the genotypes into which any given genotype can mutate become on
average more numerous than the idiotypes the genotype itself or an idiotype can
stimulate. The increase also impacts the dynamics of mutation and stimulation,
since the former becomes ever more likely than the latter. However, the meaning
of this in terms of quasispecies survival is unclear in either figure~\ref{fig4}
or figure~\ref{fig5}, despite the fact that, as noted earlier, some degree of
survival certainly occurs and seems to take shape earlier for the higher value
of $\nu$ ($0.2$ in figure~\ref{fig5}) than for the lower value ($0.05$ in
figure~\ref{fig4}).

\begin{figure}[p]
\begin{indented}
\item[]
\includegraphics[scale=0.40]{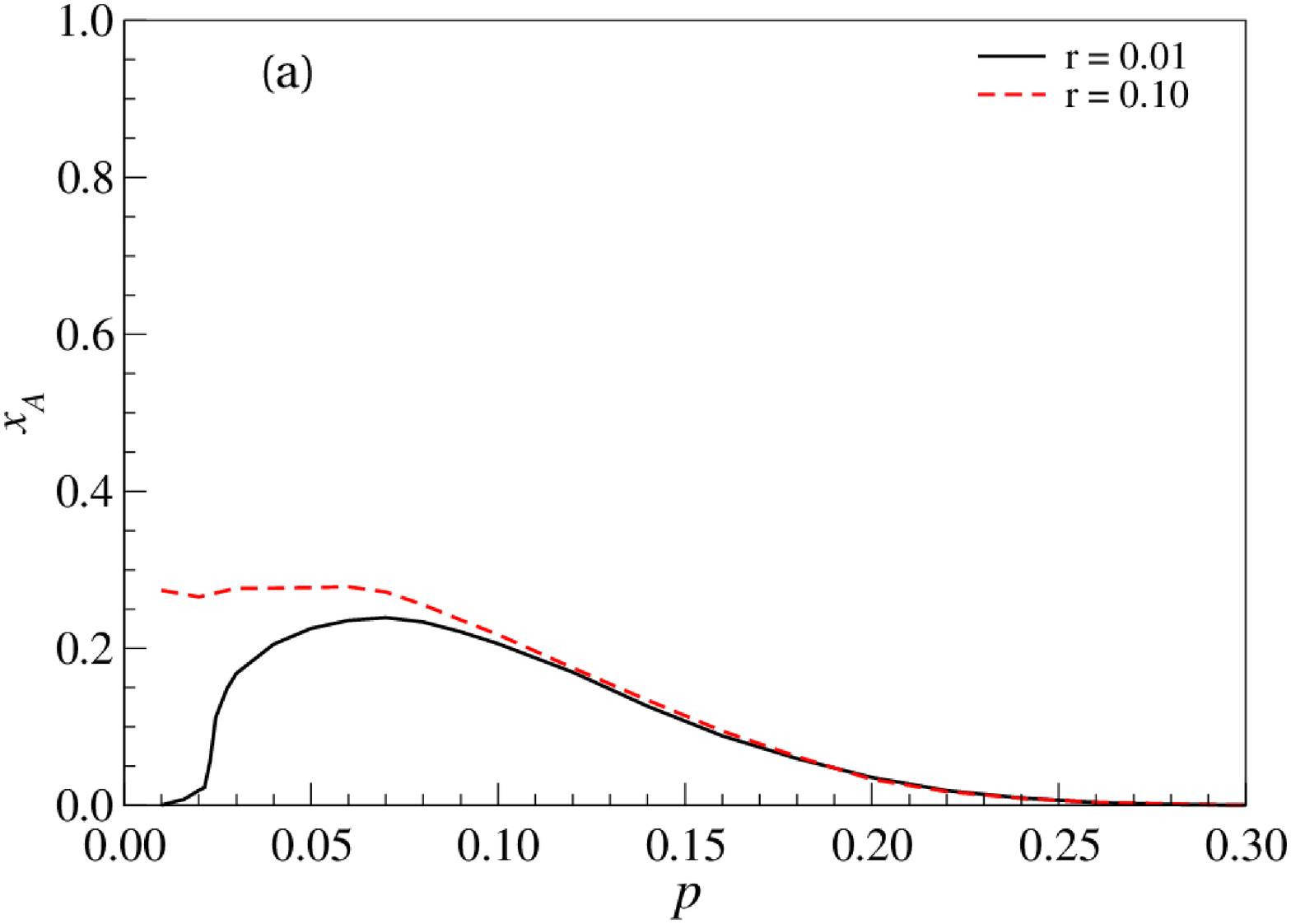}\\
\includegraphics[scale=0.40]{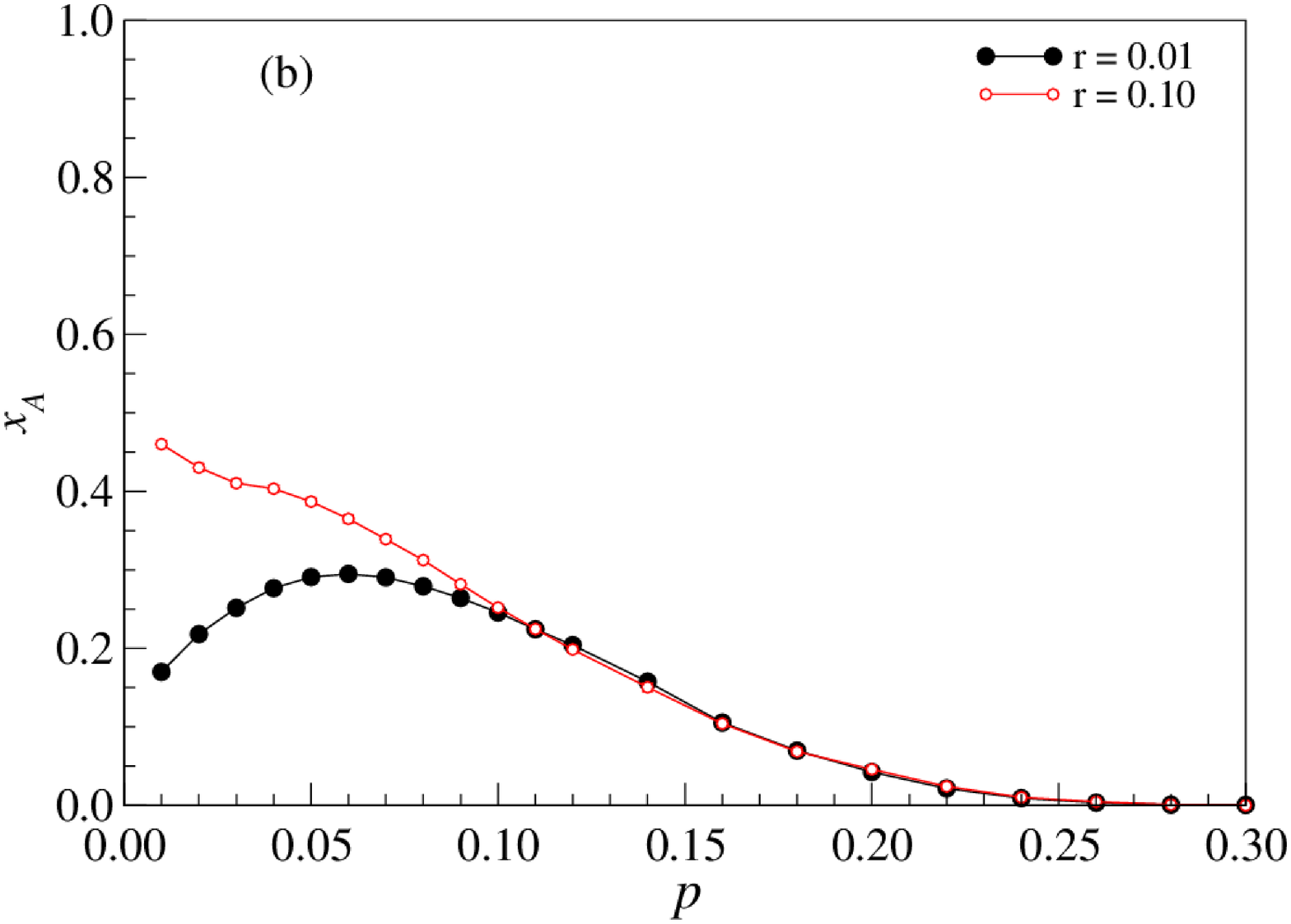}\\
\includegraphics[scale=0.40]{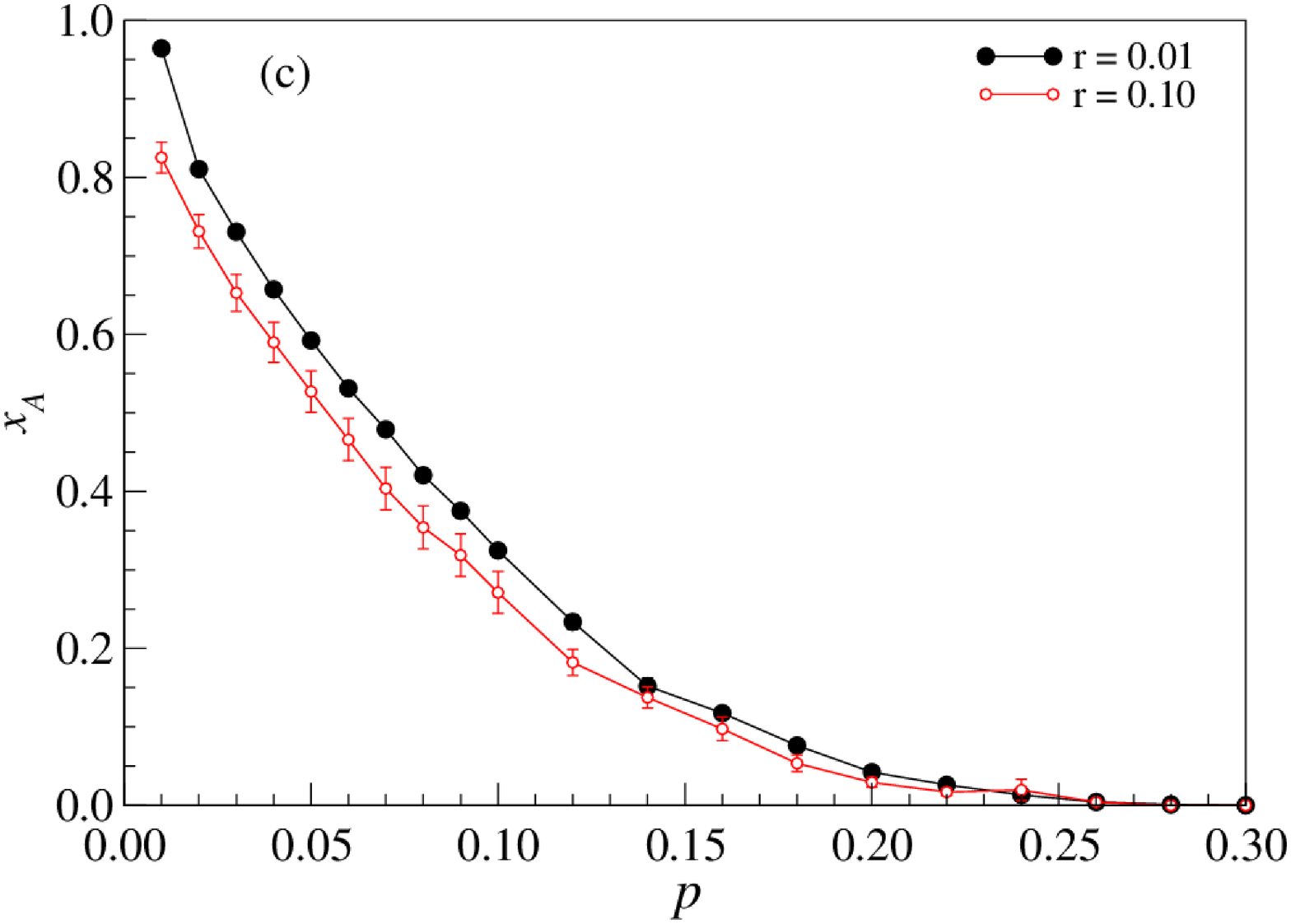}
\end{indented}
\caption{Stationary-state relative abundance of genotypes ($x_A$) as a function
of $p$. Data are given for $\lambda=\mu=0.1$ and $x_A(0)=0.1$, with $\nu=0.05$
(a), $\nu=0.1$ (b), and $\nu=0.2$ (c).}
\label{fig7}
\end{figure}

The additional data in figure~\ref{fig7} provide important further insight,
though. These data are averages of the stationary-state values of $x_A$ over the
$D$ instances used. They are given in three panels, (a) through (c),
respectively for $\nu=0.05$ (as in figures~\ref{fig3}(a) and~\ref{fig4}),
$\nu=0.1$ (as in figure~\ref{fig3}(b)), and $\nu=0.2$ (as in
figures~\ref{fig3}(c) and~\ref{fig5}). Thus, as in figure~\ref{fig3}, moving
through the panels from (a) to (c) lets the $\nu/\lambda$ ratio double at each
step, with $\nu=\lambda$ in panel (b). Each panel contains two plots against
$p$, each plot for a different value of $r$ ($0.01$ and $0.1$). All three
combinations of $p$ and $r$ values to which the panels of
figures~\ref{fig3}--\ref{fig5} correspond are present, as well as several
others. In all three panels of figure~\ref{fig7} it is clear that increasing $p$
beyond about $0.08$ progressively leads to regimes in which it becomes ever
harder for the quasispecies to survive. This, we note, is fully compatible with
all theories of the quasispecies, including our own network-based theory in
\cite{bds12}. The quasispecies can also fail at the other extreme, that in which
$p$ is made very low (e.g., $p=0.01$), but only insofar as $\nu<\lambda$ and
provided, additionally, that the value of $r$ is substantially higher than that
of $p$.

This issue of very small $p$ relative to $r$ was already addressed in
\cite{bds15} and, in the specific case of $p=0.01$ with $r=0.1$, the wild type
was found to, on average, have very restricted escape routes through mutation to
evade the action of the idiotypes, a situation that ends almost always in the
wild type's own dilution and hence the destruction of the quasispecies. As noted
in \cite{bds15}, this is to be contrasted with the case of the stand-alone
quasispecies network of \cite{bds12}, in which arbitrarily low values of $p$ are
practically a guarantee of quasispecies survival. In light of the additional
information given in figure~\ref{fig7}, what figures~\ref{fig4} and~\ref{fig5}
seem to be indicating is that the survival of the quasispecies becomes again
possible for very low values of $p$ when $\nu>0$, with $\nu=\lambda$ working as
a threshold for the value of $r$ to be at all relevant. That is, if the idiotype
population can be depleted sufficiently strongly by the action of genotypes,
then lowering $p$ substantially is no impediment to quasispecies survival.

We now turn to the issue of pathogenic idiotypes, that is, idiotypes whose
sequence representation in our model is identical to that of a genotype of high
fitness (the wild type or some other genotype whose Hamming distance to it is
very small). We do so by first defining, for $h\in\{0,1,\ldots,L\}$, the set
$B(h)$ comprising those idiotypes whose Hamming distance to the wild type is
$h$. Clearly, $\vert B(h)\vert=\vert B(L-h)\vert={L\choose h}$, where we use
$\vert X\vert$ to denote the cardinality of set $X$. We study the rise of
pathogenic idiotypes through $x_{B(h)}$, which we define to be the average
relative abundance of all idiotypes in $B(h)$. That is,
\begin{equation}
x_{B(h)}={L\choose h}^{-1}\sum_{i\in B(h)}x_i.
\end{equation} 

\begin{figure}[p]
\begin{indented}
\item[]
\includegraphics[scale=0.40]{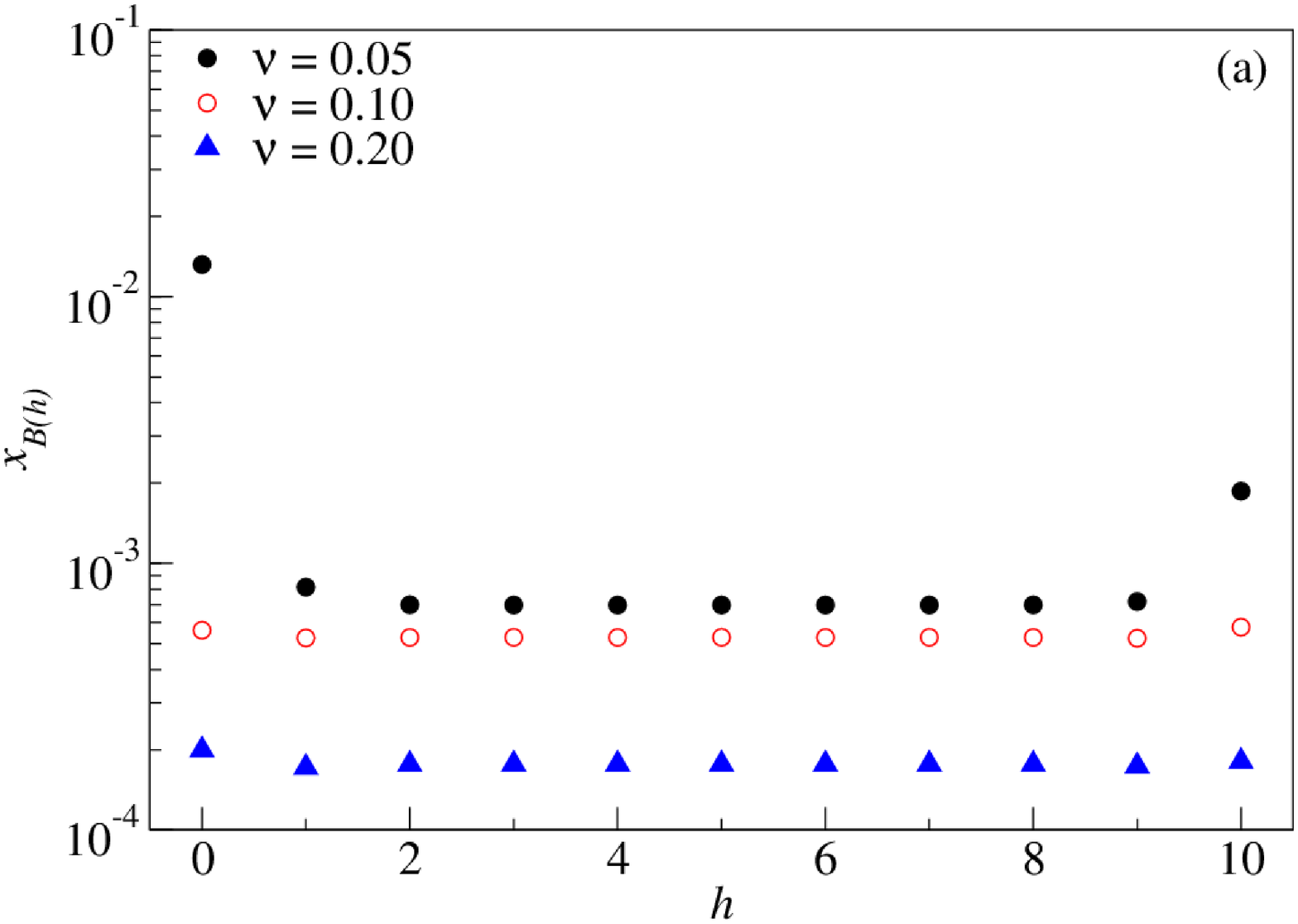}\\
\includegraphics[scale=0.40]{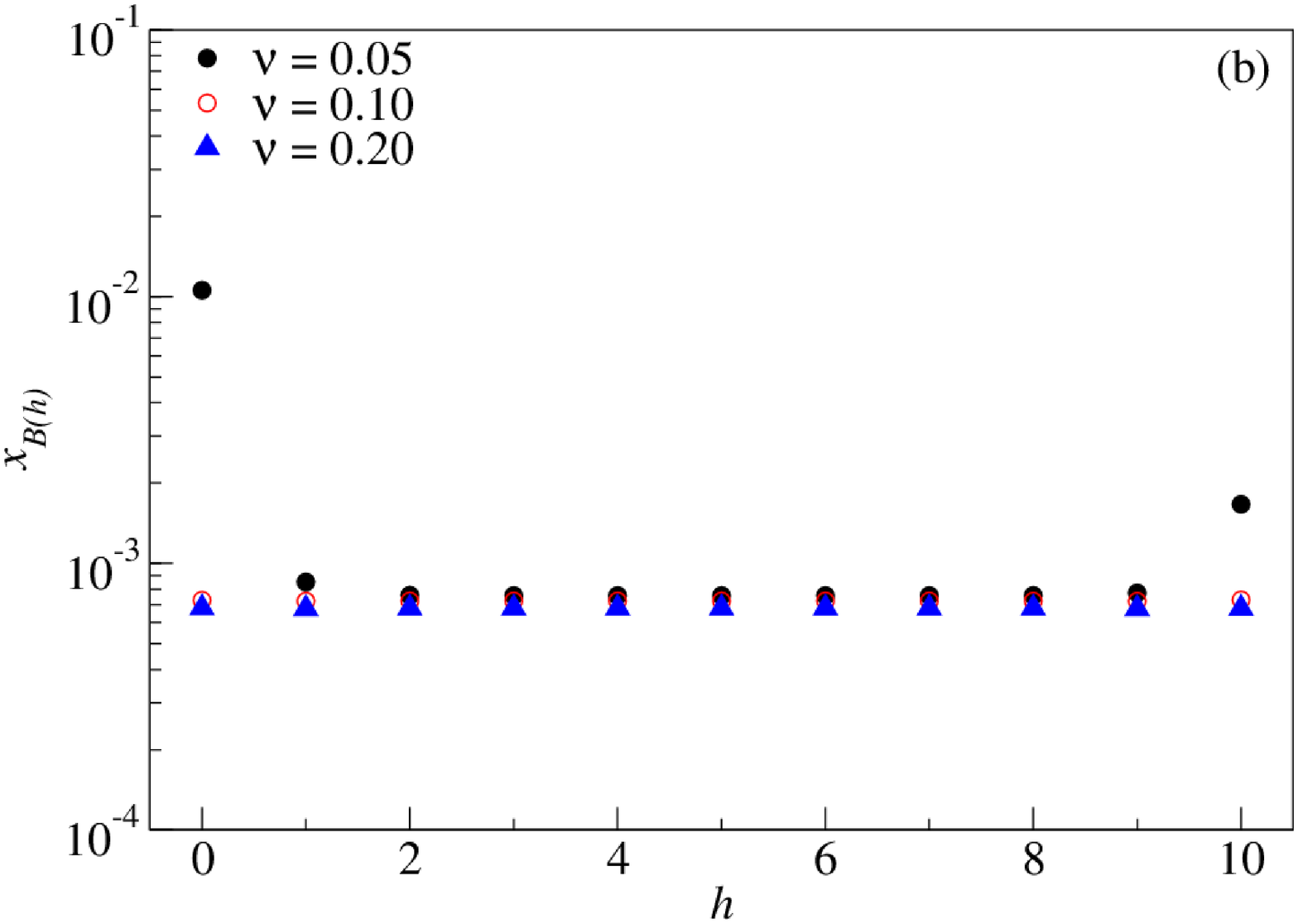}\\
\includegraphics[scale=0.40]{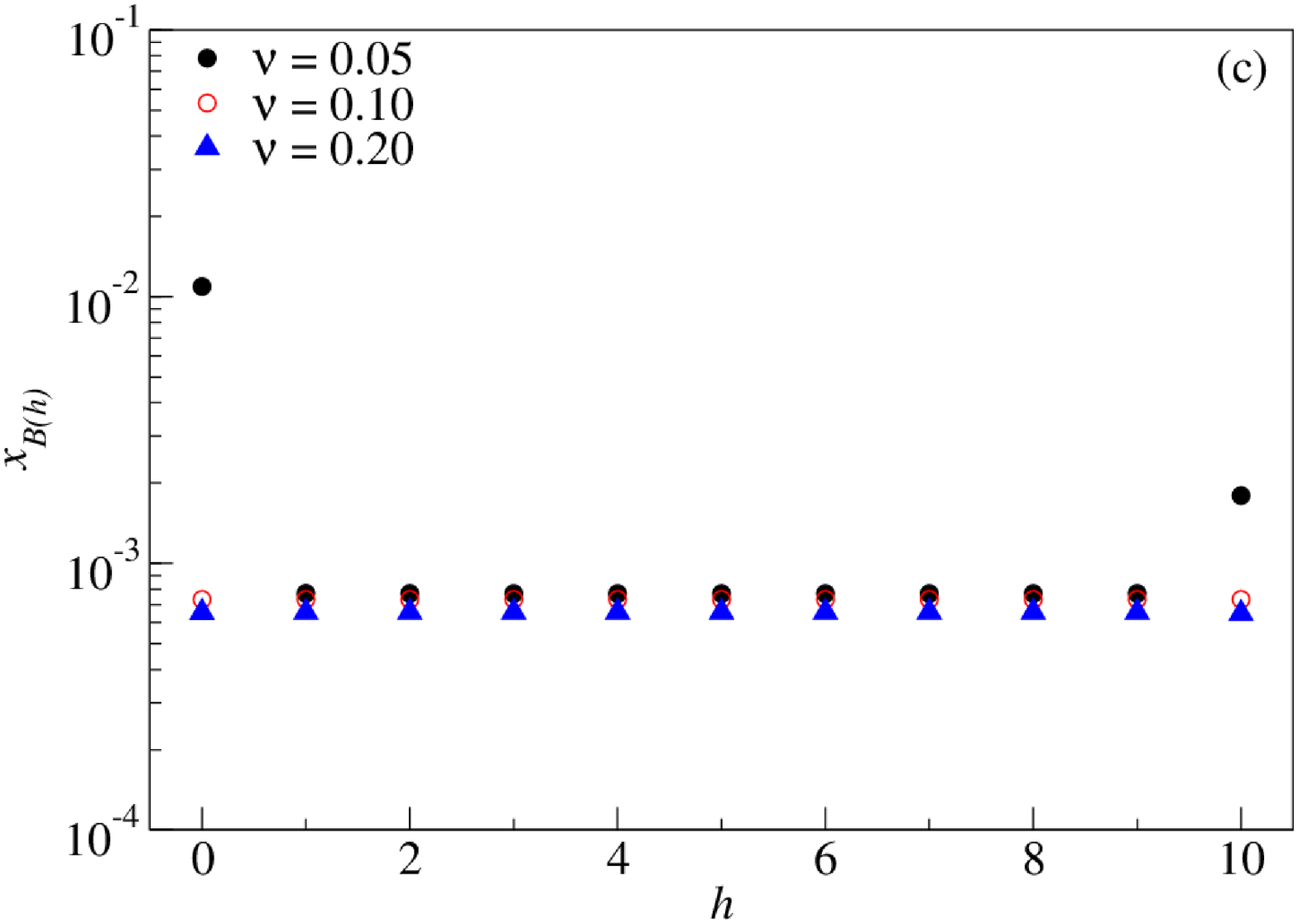}
\end{indented}
\caption{Stationary-state relative abundance of the idiotypes in $B(h)$
($x_{B(h)}$) as a function of $h$. Data are given for $\lambda=\mu=0.1$ and
$x_A(0)=0.1$, with $p=0.01$, $r=0.1$ (a), $p=r=0.1$ (b), and $p=0.1$, $r=0.01$
(c).}
\label{fig8}
\end{figure}

The stationary-state value of $x_{B(h)}$ is shown in the three panels of
figure~\ref{fig8} against $h$, each for a different combination of $p$ and $r$
values, each comprising plots for three values of $\nu$ ($0.05$, $0.1$, and
$0.2$). Further parameter values for this figure are $\lambda=\mu=0.1$ and
$x_A(0)=0.1$. As in figures~\ref{fig4} and~\ref{fig5}, the values of $p$ and $r$
are taken from the set $\{0.01,0.1\}$ so that the ratio $p/r$ varies from $0.1$
in panel (a), to $1$ in panel (b) for $p=r=0.1$, to $10$ in panel (c).
Considering all parameters, therefore, what figure~\ref{fig8} offers is a
different perspective on some of the scenarios we considered earlier,
particularly in figure~\ref{fig7}.

In order to interpret the data in figure~\ref{fig8}, first recall from
equation~(\ref{eq:xB}) that, whenever $\nu=\lambda$, and aside from the need to
renormalize for relative abundances, idiotype evolution happens in a way that is
decoupled from the dynamics of genotype mutation. When this is the case, every
idiotype receives on average the same amount of stimulation
(cf.\ equation~(\ref{eq:XB}) as well), so we expect it to contribute to the
stationary-state $x_B$ as much as any other idiotype, that is, uniformly. We
then expect $x_{B(h)}=(1-x_A)/2^L$ in the stationary state, regardless of the
value of $h$ or any of the dynamics-related parameters. All three panels in
figure~\ref{fig8} contain a $\nu=\lambda=0.1$ plot that is indeed nearly flat.
The value of $x_{B(h)}$ at which this happens can be estimated with the help of
figure~\ref{fig7}(b), where we find $x_A\approx 0.46$ for $p=0.01$ and $r=0.1$,
and $x_A\approx 0.25$ for $p=0.1$ with both $r=0.1$ and $r=0.01$. These yield
$x_{B(h)}\approx 5.3\times 10^{-4}$ (which agrees with panel (a) of
figure~\ref{fig8}) and $x_{B(h)}\approx 7.3\times 10^{-4}$ (which agrees with
panels (b) and (c)).

A similar situation of flatness with respect to $h$ occurs also for $\nu=0.2$,
particularly so in panels (b) and (c) of the figure, those in which $p\ge r$.
Once again, in all three cases this nearly flat stationary-state $x_B(h)$ occurs
at about the level given by $(1-x_A)/2^L$, where $x_A$ is the stationary-state
relative abundance of genotypes read off figure~\ref{fig7}(c) for the
appropriate combination of $p$ and $r$ values. Clearly, this nearly flat
behavior for $\nu>\lambda$ is the result of the strong pull exerted by genotypes
on idiotypes, which tends to deplete the relative abundances of the latter.
Moving to the other side of the $\nu=\lambda$ threshold reveals a different
situation, though. This is exemplified in figure~\ref{fig8} for $\nu=0.05$,
where idiotypes mimicking the wild type (i.e., with $h=0$) do appear in
non-negligible concentrations regardless of how $p$ and $r$ relate to each
other. This happens for the idiotype that is fully complementary to the wild
type as well (i.e., with $h=L$), but at a much lower concentration. That a
genotype as fit as the wild type should find itself mimicked at a significant
level of relative abundance amid the idiotypes when they are threatened with
destruction by the genotypes only mildly (i.e., for $\nu<\lambda$) indicates
clearly that our model is rich enough to give rise to the appearance of the
so-called pathogenic idiotypes.

\begin{figure}[t]
\begin{indented}
\item[]
\includegraphics[scale=0.45]{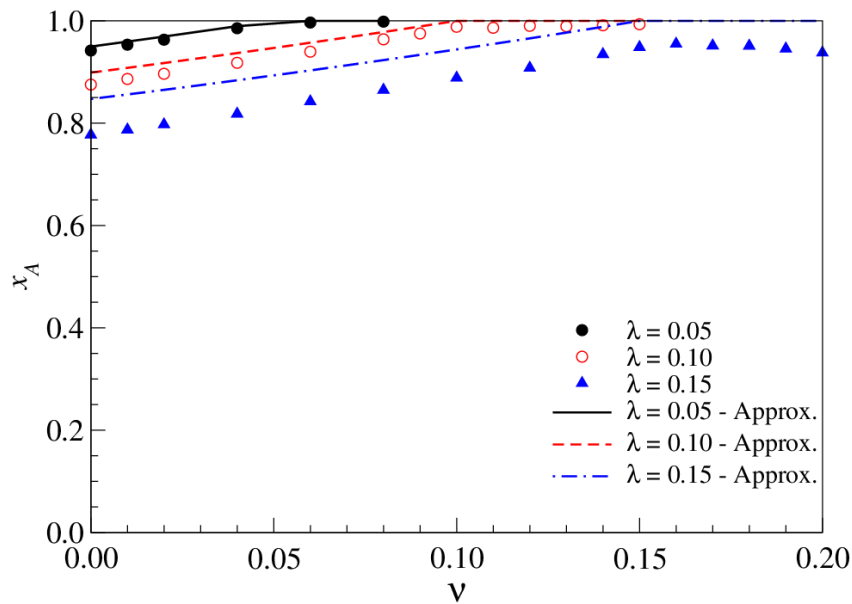}
\end{indented}
\caption{Stationary-state relative abundance of genotypes ($x_A$) as a function
of $\nu$. Data are given for $p=r=0.1$, $\mu=0.1$, and $x_A(0)=0.14$. Lines
correspond to the analytical results of section~\ref{sec:model-special}, here
used as approximations as per the formula for $x_A^+$ in
equation~(\ref{eq:roots}).}
\label{fig9}
\end{figure}

We conclude the section with an examination of figure~\ref{fig9}, where we look
at the special-case analytical results of section~\ref{sec:model-special} as
possible approximations of the general case. When we initially considered this
possibility we found out that such could only be the case if the quasispecies
were guaranteed to almost surely survive as well as be markedly more abundant
than the idiotypes. We thus concentrate on the case of $x_A(0)=0.14$, with
support from figure~\ref{fig2}. Fixing $p=r=0.1$ and $\mu=0.1$ as well,
figure~\ref{fig9} suggests that the formula for $x_A^+$ in
equation~(\ref{eq:roots}) works reasonably well as an approximation of the
stationary-state $x_A$ for a wide range of $\lambda$ and $\nu$ values (all of
which lead to $x_A(0)>x_A^-$, so $x_A=x_A^+$ really is the expected limit in the
special case, as discussed in section~\ref{sec:model-special}). This is
especially true of the lower values of $\lambda$, but is in any case remarkable,
since that equation is predicated upon $p=r=1$ as well as a flat fitness
landscape across all genotypes.

\section{Conclusion}
\label{sec:concl}

The genotype-idiotype interaction model analyzed in this paper generalizes our
previous model in \cite{bds15} through the rate parameter $\nu$, which accounts
for the possibility of idiotype removal by genotypes in the manner of
retroviruses. The relationship between $\nu$ and $\lambda$, the rate of idiotype
stimulation by both genotype and idiotype action, is crucial for the
understanding of how the two populations can be expected to behave. Setting
$\nu=\lambda$, in particular, makes the idiotype population essentially
independent of what may be happening on the genotype side. It also acts as a
sort of threshold with respect to which the quasispecies can be expected to
survive, more or less intensely, or even be destroyed.

For $\nu<\lambda$ but still nonzero, we have found that the quasispecies can
survive at a moderate level of relative abundance even for very small values of
$p$. This is in stark contrast with the case of $\nu=0$ studied in \cite{bds15},
where we found that $p$ values below a certain minimum imply the destruction of
the quasipecies. So, in a way, setting $\nu$ to a value between $0$ and
$\lambda$ seems to restore the survival abilities of the quasispecies to what
they were in the absence of the idiotypic network \cite{bds12}. Moving toward
the other end of the spectrum, with $\nu>\lambda$, affects quasispecies survival
similarly, but now at high levels of relative abundance.

In addition to the issue of quasispecies survival (with the idiotype population
following the complementary trend, either toward survival or destruction), we
have found that the wild type, the fittest of all genotypes, can find itself
mimicked amid the idiotypes more than any other genotype. It then seems that our
model is capable of capturing some of the fundamental mechanisms underlying the
appearance of pathogenic idiotypes. In fact, we have found that this holds only
for $\nu<\lambda$ (with $\nu\ge\lambda$ implying the dilution of the idiotype
population nearly uniformly across all idiotypes). This, interestingly, is in
good agreement with the discovery that autoimmune diseases can occur in HIV
patients, but mainly after the HIV population has receded in response to early
treatment by modern therapies \cite{ilbjgbchmlf14}. That is, in terms of our
model, mainly for those cases in which $\nu<\lambda$ and the genotype
population, though possibly still exerting some pull on the idiotypes, is
incapable of hampering their survival.

Thus, while our results suggest a network-theoretic framework to explain the
appearance of pathogenic idiotypes, they also raise important further questions.
For example: should the idiotypic network be allowed to evolve without any
connections to the network of a certain quasispecies, and should it result in a
configuration particularly capable of resisting that quasispecies' wild type,
what would happen if the two networks were finally brought together? Would the
wild type still be mimicked and possibly give rise to autoimmune disease? The
relevant issue here is an alleged evolutionary trade-off, recently suggested in
\cite{vs14}, between the immune system's fitness to fight and its ability to
keep autoimmunity curbed. All our results so far, both in the present work and
in the previous one \cite{bds15}, come from studying the dynamics of the entire
network, comprising genotypes and idiotypes alike, but clearly from an
evolutionary perspective it also makes sense to study the interaction of the
network's two halves only after the idiotypic half has undergone evolution
separately while interacting with different quasispecies. We believe our model,
possibly enhanced by still further additions, may be able to provide some useful
insight in such a scenario as well.

\ack
We acknowledge partial support from CNPq, CAPES, and a FAPERJ BBP grant.

\bibliographystyle{iopart-num}
\section*{References}
\bibliography{inet2}

\providecommand{\newblock}{}
\begin{thebibliography}{10}
\expandafter\ifx\csname url\endcsname\relax
  \def\url#1{{\tt #1}}\fi
\expandafter\ifx\csname urlprefix\endcsname\relax\def\urlprefix{URL }\fi
\providecommand{\eprint}[2][]{\url{#2}}

\bibitem{e71}
Eigen M 1971 {\em Naturwissenschaften\/} {\bf 58} 465--523

\bibitem{es77}
Eigen M and Schuster P 1977 {\em Naturwissenschaften\/} {\bf 64} 541--565

\bibitem{d09}
Domingo E 2009 {\em Contrib. Sci.\/} {\bf 5} 161--168

\bibitem{la10}
Lauring A~S and Andino R 2010 {\em PLoS Pathog.\/} {\bf 6} e1001005

\bibitem{mlcgm10}
M\'{a}s A, L\'{o}pez-Gal\'{\i}ndez C, Cacho I, G\'{o}mez J and Mart\'{\i}nez
  M~A 2010 {\em J. Mol. Biol.\/} {\bf 397} 865--877

\bibitem{bds12}
Barbosa V~C, Donangelo R and Souza S~R 2012 {\em J. Theor. Biol.\/} {\bf 312}
  114--119

\bibitem{bds15}
Barbosa V~C, Donangelo R and Souza S~R 2015 {\em J. Stat. Mech.\/}  P01022

\bibitem{j74}
Jerne N~K 1974 {\em Ann. Inst. Pasteur Imm.\/} {\bf C125} 373--389

\bibitem{b59}
Burnet F~M 1959 {\em The Clonal Selection Theory of Acquired Immunity\/}
  (Cambridge, UK: Cambridge University Press)

\bibitem{f95}
Forsdyke D~R 1995 {\em FASEB J.\/} {\bf 9} 164--166

\bibitem{fabc04}
Flores L~E, Aguilar E~J, Barbosa V~C and Carvalho L~A~V 2004 {\em J. Theor.
  Biol.\/} {\bf 229} 311--325

\bibitem{b07}
Behn U 2007 {\em Immunol. Rev.\/} {\bf 216} 142--152

\bibitem{mkbmqtcb11}
Madi A, Kenett D~Y, Bransburg-Zabary S, Merbl Y, Quintana F~J, Tauber A~I,
  Cohen I~R and Ben-Jacob E 2011 {\em PLoS ONE\/} {\bf 6} e17445

\bibitem{yfkbm15}
Yang S, Fujikado N, Kolodin D, Benoist C and Mathis D 2015 {\em Science\/} {\bf
  348} 589--594

\bibitem{p83}
Plotz P~H 1983 {\em Lancet\/} {\bf 322} 824--826

\bibitem{o87}
Oldstone M~B~A 1987 {\em Cell\/} {\bf 50} 819--820

\bibitem{sm90}
Shoenfeld Y and Mozes E 1990 {\em FASEB J.\/} {\bf 4} 2646--2651

\bibitem{pmetal15}
Penaloza-MacMaster P, Barber D~L, Wherry E~J, Provine N~M, Teigler J~E,
  Parenteau L, Blackmore S, Borducchi E~N, Larocca R~A, Yates K~B, H~Shen
  W~N~H, Sommerstein R, Pinschewer D~D and R~Ahmed D~H~B 2015 {\em Science\/}
  {\bf 347} 278--282

\bibitem{g07}
Goff S~P 2007 {\em Nat. Rev. Microbiol.\/} {\bf 5} 253--263

\bibitem{s12}
Stoye J~P 2012 {\em Nat. Rev. Microbiol.\/} {\bf 10} 395--406

\bibitem{zengetal14}
Zeng M, Hu Z, Shi X, Li X, Zhan X, Li X~D, Wang J, Choi J~H, Wang K~w,
  Purrington T, Tang M, Fina M, DeBerardinis R~J, Moresco E~M~Y, Pedersen G,
  McInerney G~M, {Karlsson Hedestam} G~B, Chen Z~J and Beutler B 2014 {\em
  Science\/} {\bf 346} 1486--1492

\bibitem{be06}
Biebricher C~K and Eigen M 2006 What is a quasispecies? {\em Quasispecies:
  Concept and Implications for Virology\/} ({\em Current Topics in Microbiology
  and Immunology\/} vol 299) ed Domingo E (Berlin, Germany: Springer) pp 1--31

\bibitem{ilbjgbchmlf14}
Iordache L, Launay O, Bouchaud O, Jeantils V, Goujard C, Boue F, Cacoub P,
  Hanslik T, Mahr A, Lambotte O, Fain O and {associated authors} 2014 {\em
  Autoimmun. Rev.\/} {\bf 13} 850--857

\bibitem{vs14}
Volkman H~E and Stetson D~B 2014 {\em Nat. Immunol.\/} {\bf 15} 415--422

\end{thebibliography}

\end{document}